\documentclass[aps,pra,twocolumn,superscriptaddress,floatfix]{revtex4-1}
\usepackage[a4paper,textwidth=183 mm,textheight = 247mm]{geometry}
\usepackage{mathptmx}
\usepackage{graphicx}
\usepackage[tbtags]{amsmath}
\usepackage{amsthm}
\usepackage{empheq}
\usepackage{amssymb}
\usepackage{mathrsfs}
\usepackage{float}
\usepackage{array}

\usepackage[sort&compress]{natbib}
\setcitestyle{super,open={},close={},citesep={,}}

\renewcommand{\figurename}{\textbf{Figure}}

\usepackage{caption}
\DeclareCaptionLabelSeparator{line}{\,\textbf{\textbar}\,}
\usepackage{color}
\usepackage{dcolumn}
\usepackage{bm}
\usepackage{here}
\usepackage{braket}
\allowdisplaybreaks

\newcommand{\mean}[1]{\langle#1\rangle}
\newcommand{\abs}[1]{\left\vert#1\right\vert}

\begin{document}

\title{Time-Domain Multiplexed 2-Dimensional Cluster State:\\ Universal Quantum Computing Platform}
\author{Warit Asavanant} \affiliation{Department of Applied Physics, School of Engineering, The University of Tokyo, 7-3-1 Hongo, Bunkyo-ku, Tokyo 113-8656, Japan}

\author{Yu Shiozawa} \affiliation{Department of Applied Physics, School of Engineering, The University of Tokyo, 7-3-1 Hongo, Bunkyo-ku, Tokyo 113-8656, Japan}

\author{Shota Yokoyama} \affiliation{Centre for Quantum Computation and Communication Technology, School of Engineering and Information Technology, University of New South Wales Canberra, ACT 2600, Australia}

\author{Baramee Charoensombutamon} \affiliation{Department of Applied Physics, School of Engineering, The University of Tokyo, 7-3-1 Hongo, Bunkyo-ku, Tokyo 113-8656, Japan}

\author{Hiroki Emura} \affiliation{Department of Applied Physics, School of Engineering, The University of Tokyo, 7-3-1 Hongo, Bunkyo-ku, Tokyo 113-8656, Japan}

\author{Rafael N. Alexander}\affiliation{Center for Quantum Information and Control, University of New Mexico, MSC07-4220, Albuquerque, New Mexico 87131-0001, USA}

\author{Shuntaro Takeda} \affiliation{Department of Applied Physics, School of Engineering, The University of Tokyo, 7-3-1 Hongo, Bunkyo-ku, Tokyo 113-8656, Japan}

\author{Jun-ichi Yoshikawa} \affiliation{Department of Applied Physics, School of Engineering, The University of Tokyo, 7-3-1 Hongo, Bunkyo-ku, Tokyo 113-8656, Japan}

\author{Nicolas C. Menicucci}\affiliation{Centre for Quantum Computation and Communication Technology, School of Science, RMIT University, Melbourne, Victoria 3001, Australia}

\author{Hidehiro Yonezawa}\affiliation{Centre for Quantum Computation and Communication Technology, School of Engineering and Information Technology, University of New South Wales Canberra, ACT 2600, Australia}

\author{Akira Furusawa} \email{akiraf@ap.t.u-tokyo.ac.jp}\affiliation{Department of Applied Physics, School of Engineering, The University of Tokyo, 7-3-1 Hongo, Bunkyo-ku, Tokyo 113-8656, Japan}

\maketitle
\textbf{Quantum computation promises applications \cite{Shor1997,Lloyd1997,Nielsen2000} that are thought to be impossible with classical computation. To realize practical quantum computation, the following three properties will be necessary: universality, scalability, and fault-tolerance. Universality is the ability to execute arbitrary multi-input quantum algorithms. Scalability means that computational resources such as logical qubits can be increased without requiring exponential increase in physical resources. Lastly, fault-tolerance is the ability to perform quantum algorithms in presence of imperfections and noise. A promising approach to scalability was demonstrated with the generation of one-million-mode 1-dimensional cluster state \cite{Yoshikawa2016}, a resource for one-input computation in measurement-based quantum computation (MBQC) \cite{Raussendorf2001,Menicucci2006}. The demonstration was based on time-domain multiplexing (TDM) approach \cite{Menicucci2011,Yokoyama2013,Alexander2018} using continuous-variable (CV) optical flying qumodes (CV analogue of qubit). Demonstrating universality, however, has been a challenging task for any physical system and approach. Here, we present, for the first time among any physical system, experimental realization of a scalable resource state for universal MBQC: a 2-dimensional cluster state. We also prove the universality and give the methodology for utilizing this state in MBQC. Our state is based on TDM approach that allows unlimited resource generation regardless of the coherence time of the system. As a demonstration of our method, we generate and verify a 2-dimensional cluster state capable of about 5,000 operation steps on 5 inputs. Note that there are no limitation on the number of operation steps, and the number of inputs can be increased by several orders of magnitude with current technology \cite{Ast2013,Aasi2013}. Furthermore, by increasing the squeezing level, fault-tolerance can also be achieved by combining our method with error correction schemes for CV cluster states \cite{Menicucci2014,Fukui2018}. Thus, our work not only achieves scalable generation of resource for universal MBQC, but it also takes us closer to the realization of scalable fault-tolerant universal quantum computation suitable for actual applications.}

One of the many approaches to realizing a scalable fault-tolerant universal quantum computer \cite{Nielsen2000,Raussendorf2001,Menicucci2006,Tameem2018} is MBQC \cite{Raussendorf2001,Menicucci2006}, a paradigm where computations are implemented by an adaptive sequence of measurements on a universal multipartite entangled resource state, a so-called \textit{cluster state}, and classical feed-forward. The measurement-based model comes with many appealing features; cluster states with suitable structures can be used to implement arbitrary operations meaning that generation of a single well-designed state is sufficient for universal quantum computation. Also, measurements are often much easier to implement, control, and program compared to direct implementation of quantum gates.  These properties make MBQC a promising candidate for quantum computation.

\begin{figure*}[!htb]
\centering
\includegraphics[width=\textwidth]{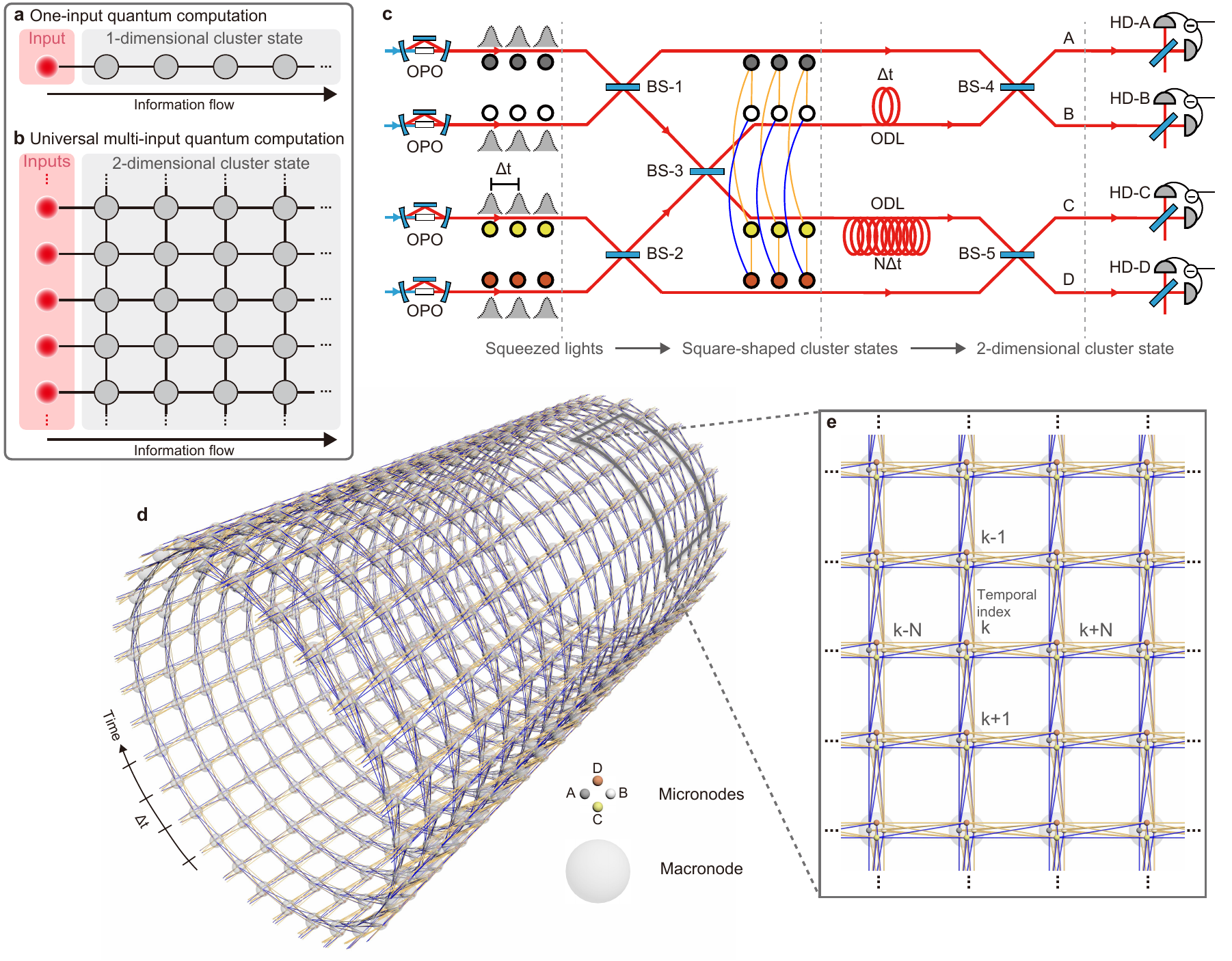}
\caption{\textbf{MBQC and 2-dimensional cluster state.} Abstract illustration of MBQC: \textbf{a,} one-input MBQC using 1-dimensional cluster state and \textbf{b,} universal multi-input MBQC using 2-dimensional cluster state. Each colored circle represents a mode, while each link represents quantum entanglement. \textbf{c,} Schematic diagram of our experimental setup for the 2-dimensional cluster state. OPO: Optical parametric oscillator. BS: Beam splitter. ODL: Optical delay line. $\Delta t$: Time interval between adjacent wave packets. $N$: Number of inputs. HD: Homodyne detector. All beam splitters are 50:50.  2-dimensional cluster state: \textbf{d,} example for the case where $N=30$; \textbf{e,} zoom in of the state. The representations of states makes use of the simplified graphical calculus \cite{Menicucci2011}. Each small node (small colored sphere) of the graph, which we call a micronode, represents a localized wave packet at each temporal index. The colors of micronodes indicate their spatial indices. Four micronodes at each temporal index $k$ can be grouped into a larger node (large gray sphere) called a macronode. The links and their colors represent how micronodes are entangled. The 2-dimensional cluster state has a helical graph structure with $N$ macronodes on every single turn of the helix. The macronode at temporal index $k$ is connected to the macronodes at $k-N$, $k-1$, $k+1$, and $k+N$. For actual experimental demonstration, we use $N=5$. Full descriptions are given in supplementary material.}
\label{fig:setupgraph}
\end{figure*}

The computational capability of MBQC is determined by the structure of the cluster state, which can be broken down into two aspects: dimension and number of modes. Cluster states with 1 dimension, i.e. where modes are entangled in a 1-dimensional chain (Fig.~\ref{fig:setupgraph}a) are resources for only one-input operations. Universality can be achieved with 2-dimensional cluster states \cite{Raussendorf2001}, i.e. where modes are entangled in 2-dimensional lattice fashion (Fig.~\ref{fig:setupgraph}b). The number of modes determines the number of the operations, hence the scalability. Therefore, large-scale 2-dimensional cluster state generation is essential for realizing scalable universal MBQC. 

\begin{figure*}[!htb]
\centering
\includegraphics[width=\textwidth]{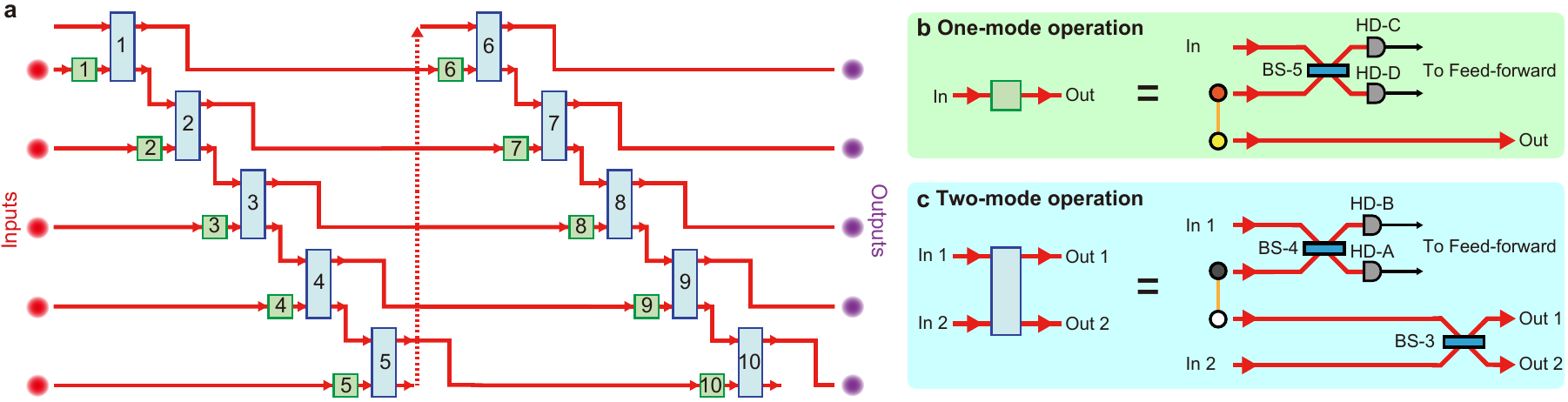}
\caption{\textbf{Quantum computation with 2-dimensional cluster state.} \textbf{a,} Equivalent quantum circuit that is implemented when our state is used. We show the case for 5 inputs where 40 light modes (10 temporal mode indices) of 2-dimensional cluster states are used. The number in each box is the index of the measured temporal modes. The circuit is composed of multiple quantum teleporters: \textbf{b,} one-mode operation and \textbf{c,} two-mode operation. Two-mode operations can be turned off by selecting the same measurement basis for HD-A and HD-B. Arbitrary multi-mode Gaussian operations can be implemented with only homodyne measurements. Non-Gaussian operations are implemented via non-Gaussian measurements. Classical feed-forward does not have to be implemented immediately after homodyne measurements and can be delayed to the end of the computation for Gaussian-only computations.}
\label{fig:computation}
\end{figure*}   

Despite its importance and many proposals on a variety of physical systems \cite{Menicucci2011,Yang2016,Alexander2016,Alexander2018,Grimsmo2017,Rudolph2017,Wunderlich2009,Albarran2018,Mamaev2018}, 2-dimensional cluster states have not been realized up until now due to their complexity. For stationary qubits \cite{Rudolph2017,Wunderlich2009,Albarran2018,Mamaev2018}, such as superconducting qubits and ion-trap qubits, we would have to prepare and spatially arrange a large number of qubits. This greatly increases the technological demands and experimental complexity as the number of qubits and the dimension of the cluster state increases. On the other hand, flying qubits (qumodes) overcome such difficulties. For CV optical systems, a flying qumode possesses rich degrees of freedom and the ability to deterministically generate entanglement. There are many schemes for scalable generation of cluster states with constant physical resources \cite{Menicucci2011,Yang2016,Alexander2016,Alexander2018,Grimsmo2017}. Among them, in the TDM approach \cite{Menicucci2011,Alexander2018}, we utilize temporally localized wave packets as modes of quantum states allowing the generation of large-scale cluster states. The scalability of the TDM approach has already been demonstrated with the generation of a one-million-mode 1-dimensional TDM cluster state \cite{Yoshikawa2016}. 

In this paper, for the first time among all physical systems, we generate a large-scale 2-dimensional cluster state for universal MBQC with a CV optical system based on the TDM approach. In our method, a 2-dimensional cluster state is continuously generated and immediately measured. Therefore, the length of the cluster state can be arbitrarily large, even with finite coherence time of the light source. This makes our method not only universal, but also highly scalable. With higher squeezing levels, fault-tolerance can be achieved by combining 2-dimensional cluster state with GKP-based error correction scheme for CV cluster states \cite{Gottesman2001,Menicucci2014,Fukui2018,Alexander2018}. We newly design and build the setup for generating a cluster state capable of universal MBQC with 5 input modes. With this setup we generate and verify 2-dimensional cluster states capable of quantum computation with about 5,000 operation steps on 5 input modes. In addition, we also prove the universality of our state and present the methodology for implementing universal MBQC with it. 

Figure~\ref{fig:setupgraph}c shows the schematic diagram for the experimental setup. While there already exist some theoretical proposals \cite{Menicucci2011,Alexander2018}, this setup was newly designed to improve experimental feasibility. To generate a 2-dimensional TDM cluster state, we first generate multiple temporally localized square-shaped cluster states on four spatially separated beams. To make the 2-dimensional structure, two optical delay lines are used. The time delay of one of the delay lines, $\Delta t$, is equal to the time interval between temporal modes of the state, while time delay of the longer delay line is equal to $N\Delta t$, where $N$ is the number of the input modes for quantum computation. After delaying modes on two beams, we connect them to temporal modes on two nondelayed beams and generate the 2-dimensional cluster state. Figure~\ref{fig:setupgraph}d shows the entanglement structure of 2-dimensional cluster state that could be generated with this scheme for the case where $N=30$. From Fig.~\ref{fig:setupgraph}e, we can see that although the temporal modes (micronodes) are connected in a complex way, if we group the micronodes with the same temporal mode indices into a larger node (macronode), the macronodes are indeed connected in 2-dimensional fashion. With four squeezed light sources, five beam splitters, and two optical delay lines, our generation method provides a simple yet scalable generation method of arbitrarily large 2-dimensional cluster states. Note that the input states can be easily inserted into the 2-dimensional cluster state via an optical switch after the longer optical delay line \cite{Yokoyama2013, Alexander2018}. For experimental demonstration, we pick $N=5$ and $\Delta t=40$ ns. $N$ can be increased by increasing the length of the longer optical delay line and the physical limitation of $N$ is that $N\Delta t$ must be below the coherence time of the system.

\begin{figure*}[!htb]
\centering
\includegraphics[width=\textwidth]{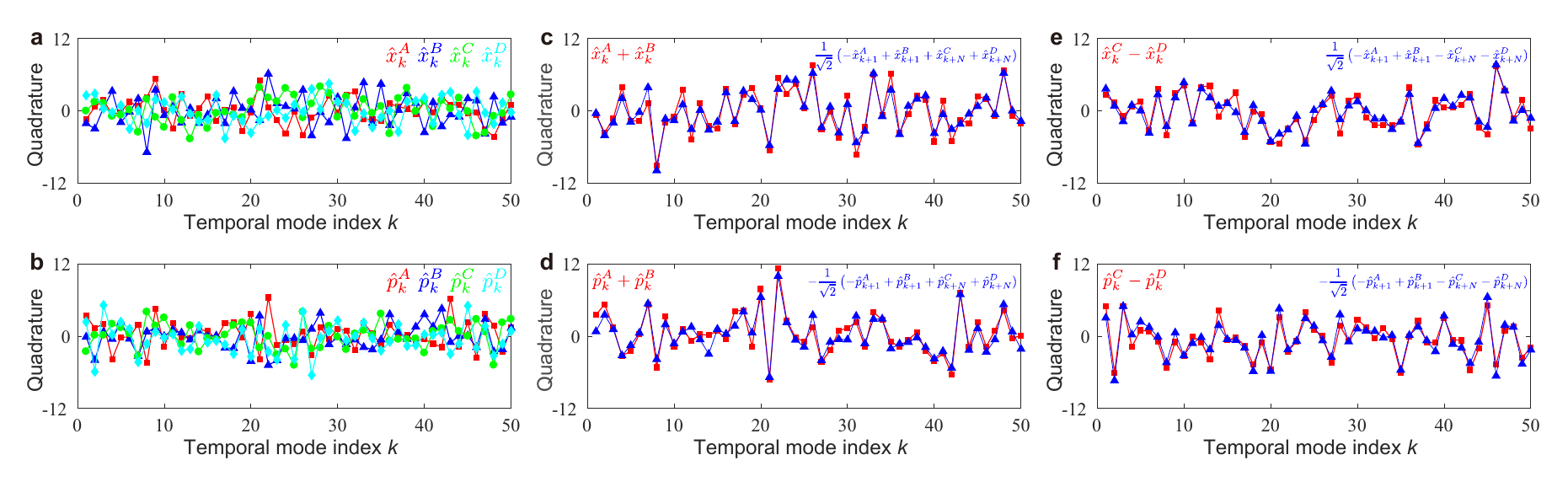}
\caption{\textbf{Quadrature values and four types of quadrature correlations of the first 50 temporal mode indices.} \textbf{a,b,} Single shot quadrature values of $\hat{x}_{k}^{j}$ and $\hat{p}_{k}^{j}$ obtained by processing time-domain signals from homodyne detector-$j$ ($j=A,B,C,D$). \textbf{c-f,} Correlations of quadrature values corresponding to $\hat{\delta}_{k}^{(x,1)}$, $\hat{\delta}_{k}^{(p,1)}$, $\hat{\delta}_{k}^{(x,2)}$, and $\hat{\delta}_{k}^{(p,2)}$, respectively. While the quadrature values measured at each homodyne detector seem to be just fluctuating randomly around zero, we observe four types of strong quantum correlations between six quadrature values with different temporal mode index $k$ and spatial index $j$.}
\label{fig:quad}
\end{figure*}
\begin{figure*}[!htb]
\centering
\includegraphics[width=\textwidth]{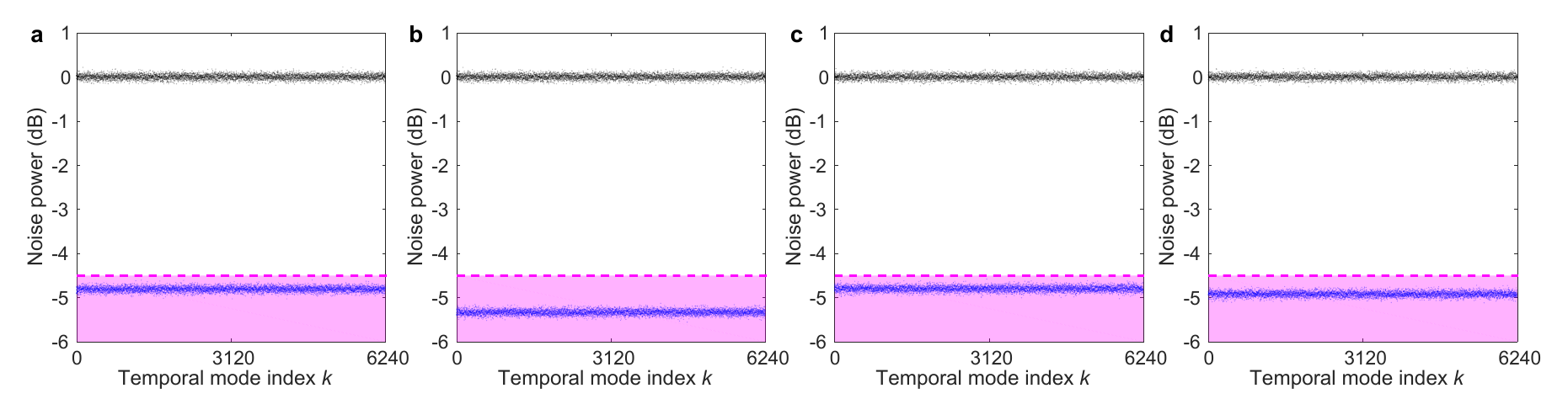}
\caption{\textbf{Verification of generated 2-dimensional cluster state for 24,960 temporal modes.} \textbf{a-d,} Measurement results for each type of nullifier: $\hat{\delta}_{k}^{(x,1)}$, $\hat{\delta}_{k}^{(p,1)}$, $\hat{\delta}_{k}^{(x,2)}$, and $\hat{\delta}_{k}^{(p,2)}$, respectively. Black points: measured variances of shot noise which are used as reference levels. Blue points: measured variances of nullifiers. Purple regions: regions where the variances of the nullifiers are below $-4.5$ dB compared to shot noise, which indicate entanglement. The variances of four types of nullifiers satisfied quantum inseparability criteria up to $k=6,240$ corresponding to $4\times 6,240=24,960$ temporal modes.}
\label{fig:nullifier}
\end{figure*} 

Next, we will describe the methodology for implementing operations with our state and prove the universality. In CV quantum computation, the ability to perform arbitrary multi-mode Gaussian operations and at least one single-mode non-Gaussian operation are required for universality \cite{Bartlett2002}.  Figure \ref{fig:computation}a shows the diagram of the equivalent quantum circuit that is implemented when this state is used for the case with $N=5$ and 10 temporal mode indices. The circuit in Fig.~\ref{fig:computation}a is composed of one-mode operations and two-mode operations based on quantum teleportation. Although quantum teleportation is usually considered as an identity operation, it is known that by selecting homodyne measurement bases besides CV-Bell measurements, Gaussian operations with two parameters can be implemented \cite{Yokoyama2013,Alexander2018}. Since arbitrary one-mode Gaussian operations require four parameters \cite{Ukai2010}, this can be implemented with two operation steps (equivalent to measurement of two macronodes). By programming the measurement basis of four homodyne detectors, we can implement arbitrary multi-mode Gaussian operations. Non-Gaussian operations can be implemented by replacing HD-C or HD-D with either a photon detector \cite{Menicucci2014} or a quantum circuitry for non-Gaussian measurement \cite{Alexander2018}. Another alternative would be realizing non-Gaussian quadrature phase gates \cite{Marek2018} on the cluster state by injecting non-Gaussian ancillary states and performing adaptive homodyne measurements. If we use GKP error correction, there is also another alternative; it was recently shown that with GKP qubits, Gaussian operations, and heterodyne measurements (which is just a combination of beam splitter and homodyne measurements), it is possible to generate distillable magic states for non-Gaussian operations \cite{Baragiola2019}. This method is extremely powerful because it reduces the necessary resource for non-Gaussian operations and fault-tolerance into only GKP qubits with sufficient purity and high enough squeezing. The detailed proofs of the universality and details on classical feed-forward are given in the supplementary material. Besides ancillary states for non-Gaussian operations, regardless of the size and input mode number $N$ of the cluster state, we only require four homodyne detectors to individually address each mode and implement computation.

To characterize the generated cluster state, we introduce nullifiers. The nullifier $\hat{\delta}$ of a quantum state $\ket{G}$ is an operator that becomes 0 when acting on state $\ket{G}$, i.e. $\hat{\delta}\ket{G}=0$. For $M$-mode Gaussian quantum states, it is sufficient to specify $M$ independent nullifiers to characterize the quantum state. For each temporal mode index $k$, the 2-dimensional cluster state in this experiment has nullifiers of the form,
\begin{align*}
\hat{\delta}_{k}^{(x,1)}&=\hat{x}_{k}^{A}+\hat{x}_{k}^{B}-\frac{1}{\sqrt{2}}\left(-\hat{x}_{k+1}^{A}+\hat{x}_{k+1}^{B}+\hat{x}_{k+N}^{C}+\hat{x}_{k+N}^{D}\right),\\
\hat{\delta}_{k}^{(p,1)}&=\hat{p}_{k}^{A}+\hat{p}_{k}^{B}+\frac{1}{\sqrt{2}}\left(-\hat{p}_{k+1}^{A}+\hat{p}_{k+1}^{B}+\hat{p}_{k+N}^{C}+\hat{p}_{k+N}^{D}\right),\\
\hat{\delta}_{k}^{(x,2)}&=\hat{x}_{k}^{C}-\hat{x}_{k}^{D}-\frac{1}{\sqrt{2}}\left(-\hat{x}_{k+1}^{A}+\hat{x}_{k+1}^{B}-\hat{x}_{k+N}^{C}-\hat{x}_{k+N}^{D}\right),\\
\hat{\delta}_{k}^{(p,2)}&=\hat{p}_{k}^{C}-\hat{p}_{k}^{D}+\frac{1}{\sqrt{2}}\left(-\hat{p}_{k+1}^{A}+\hat{p}_{k+1}^{B}-\hat{p}_{k+N}^{C}-\hat{p}_{k+N}^{D}\right),
\end{align*}
where $\hat{x}_{k}^{j}$ and $\hat{p}_{k}^{j}$ are quadrature operators at temporal mode index $k$ and at spatial index $j$, satisfying $[\hat{x}_{k}^{j},\hat{p}_{l}^{m}]=i\hbar\delta_{kl}\delta_{mj}$. State verification can be done by measuring variances of all nullifiers at each temporal mode index.

Figure \ref{fig:quad} shows the quadrature values and quantum correlations of quadratures corresponding to each type of nullifier. These quadrature values are obtained by processing the time-domain electrical signal from each homodyne detector. While the original quadrature values do not seem to possess any distinct correlations, four types of quantum correlations corresponding to four squeezed light sources are revealed when appropriate linear combinations are taken. In the ideal case, the correlations are perfect and the variances of the nullifiers become 0. In the non-ideal case, we can verify the 2-dimensional entanglement structure of the state if we observe that the variances of all the nullifiers are below $-4.5$ dB comparing to shot noise \cite{Loock2003}.

Figure \ref{fig:nullifier} shows the measurement results of four types of nullifiers at each temporal mode index $k$. The nullifiers are observed to be below $-4.5$ dB for every $k$ up to 6,240. The means of the variances for each type of nullifier are $-4.82\pm0.06$ dB, $-5.34\pm0.06$ dB, $-4.81\pm0.06$ dB, and $-4.93\pm0.06$ dB, respectively. These values are in good agreement with the experimental parameters. The statistical errors are the main contributors to the error bars, which can be arbitrarily decreased by increasing the number of events used for calculating the nullifiers. Note that there are no corrections for experimental imperfections, and the nullifiers do not degrade with the increasing $k$, suggesting that $k$ can be arbitrarily large. Since the temporal mode index number is the same as the number of macronodes and there are four micronodes at each macronode, we have verified the entanglement structure of a 2-dimensional cluster state for 5-input-mode computation, possessing 6,240 macronodes which corresponds to 24,960 micronodes. Moreover, because one micronode gives one degree of freedom for measurement in MBQC, the verified state can be translated roughly to resource capable of 5,000 operation steps on 5 input modes.

In summary, we have, for the first time in any physical system, experimentally realized a large-scale 2-dimensional cluster state capable of implementing universal CV MBQC. While our experiment is proof of principle for the 5 input modes case, with the development of gigahertz bandwidth squeezed light \cite{Ast2013} and large optical delay lines in gravitational wave detectors \cite{Aasi2013}, a number of the input modes on the order of $10^{4}$ is attainable with currently available technology. Moreover, if we consider the ultimate limitation of our method, namely, coherence time, even with modest sub-hertz light sources \cite{Wu2016}, the number of the input modes could be potentially increased to about $10^{10}$ modes. While the current squeezing level of our cluster state is around $-5$ dB and the record for the state-of-the-art squeezing level is around $-15$ dB \cite{Vahlbruch2016}, the currently known threshold for CV error correction scheme is $-20.5$ dB \cite{Menicucci2014}. Thus, improvement of squeezed light source and lowering the threshold for error correction are required to achieve fault-tolerance. The latter might be possible by extending the structure of the cluster state to 3 dimensions which allows concatenation with other quantum error correcting codes, such as the surface code \cite{Raussendorf2007}.

\section*{Method}
Optical parametric oscillators (OPO) with bandwidth of approximately 80 MHz were used as sources for squeezed vacua. Build-up cavities were used to amplify the pump power of the OPOs. The width of the wave packet was picked to be $\Delta t=40$ ns. Free space delay line of 12 m length, and optical fiber delay line of 40 m length (equivalent to about 60 m in free space) were used for $\Delta t$ and $N \Delta t$ delay line with $N=5$, respectively. For verification, homemade homodyne detectors with 100 MHz bandwidth were used and 12,000 events were used for calculating statistical quantity at each temporal mode index $k$. See supplementary material for more details on experimental setup and derivation of threshold for verification of quantum inseparability.


\section*{Acknowledgments}
This work was partly supported by JSPS KAKENHI, UTokyo Foundation, and the Australian Research Council Centre of Excellence for Quantum Computation and Communication Technology (Project No. CE170100012). W. A. acknowledges financial support from the Japan Society for the Promotion of Science (JSPS). Y. S. acknowledges financial support from the Advanced Leading Graduate Course for Photon Science (ALPS). B. C. acknowledges financial support from Program of Excellence in Photon Science (XPS). R. N. A. is supported by National Science Foundation Grant No. PHY-1630114.

\section*{Author contributions}
W. A. and Y. S. planned and designed the experiment. S. Y., S. T., J. Y., H. Y., and A. F. supervised this project. R. N. A., N. C. M., W. A., and S. Y. formulated the theory for this experiment. Y. S. and W. A. designed the actual optical system.  W. A. conceived the method for stabilizing the experimental system and perform analysis necessary for experimental realization. W. A., Y. S., and H. E. built the experimental system. W. A., Y. S., and B. C. conducted the experiment and obtained the experimental data. Y. S., B. C., and W. A. performed the data analysis. W. A. wrote the manuscript with assistance from Y. S., B. C., S. Y., R. N. A., S. T., J. Y., N. C. M., H. Y., and A. F.


\section*{Competing financial interests}
The authors declare no competing financial interests.


\makeatletter
\setcounter{section}{0}
\setcounter{figure}{0} 

\renewcommand{\thesection}{S\arabic{section}}
\renewcommand{\thesubsection}{S\arabic{section}.\arabic{subsection}}
\renewcommand{\thesubsubsection}{S\arabic{section}.\arabic{subsection}.\arabic{subsubsection}}

\def\theequation{S\arabic{equation}}
\renewcommand{\figurename}{\textbf{Fig.}}
\renewcommand{\thefigure}{S\arabic{figure}}

\renewcommand\refname{References}
\makeatother
\clearpage
\onecolumngrid

\begin{center}
{\LARGE
Supplementary Information for Time-Domain Multiplexed}\\
{\LARGE  2-Dimensional Cluster State: Universal Quantum Computing Platform}
\end{center}

\section{Experimental Setup}

\begin{figure}[ht]
\centering
\includegraphics[width=\textwidth]{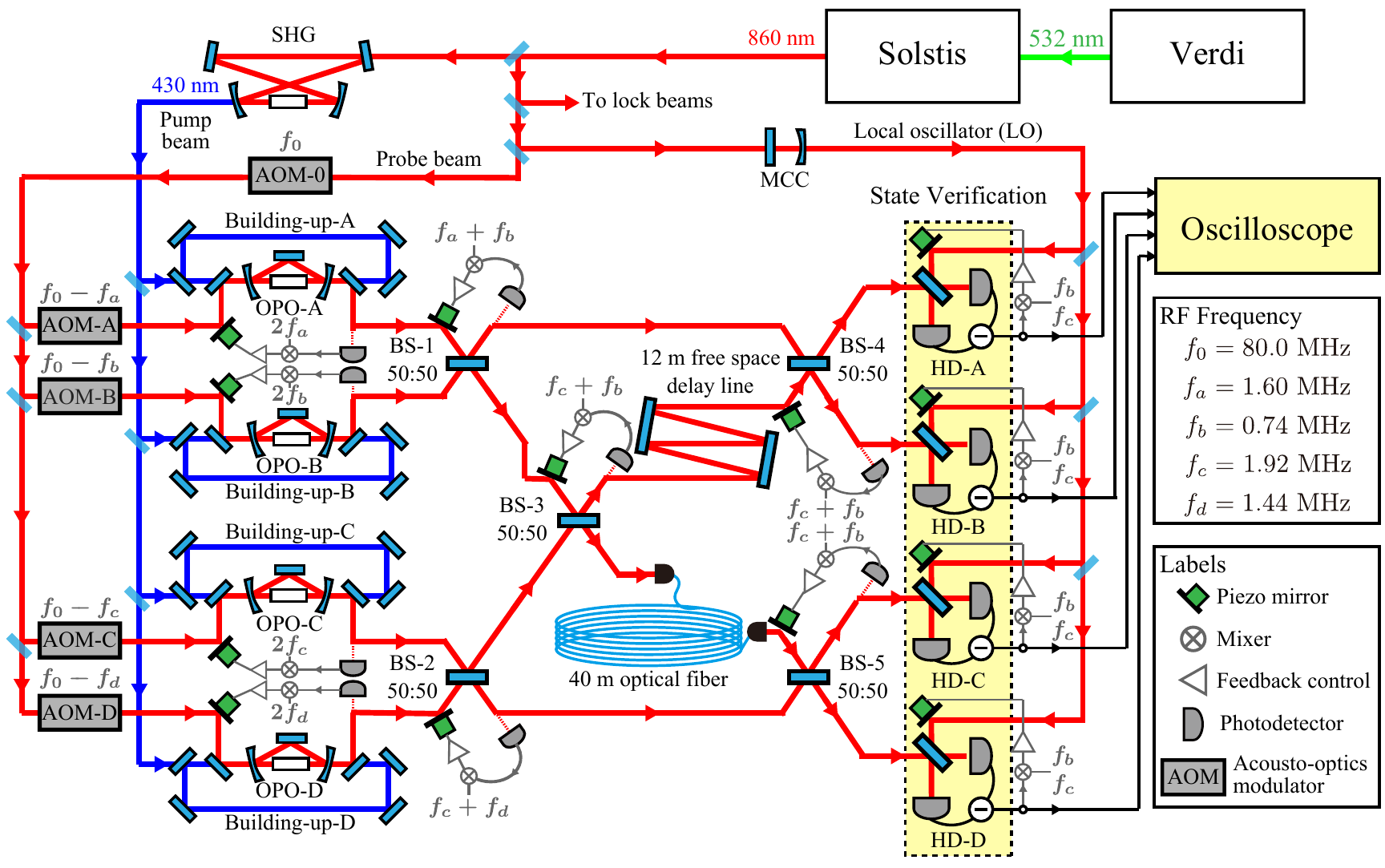}
\caption{Schematic diagram of experimental setup. OPO: Optical parametric oscillator. BS: Beam splitter. MCC: Mode cleaning cavity. SHG: Second harmonic generator. HD: Homodyne detector. Reference cavity used in matching spatial mode between SHG and OPOs is omitted and components for controlling cavity length are also omitted.}
\label{fig:expsetup}
\end{figure}

Figure \ref{fig:expsetup} shows the schematic diagram of the experimental setup. Continuous-wave (CW) Ti:Sapphire laser (SolsTis, M Squared lasers) of wavelength 860 nm is used as the light source for this experiment. The output of the laser is about 1.4 W, and is divided into mainly three parts; Around 1W is used for pumping a second harmonic generator (SHG) to generate a 430 nm pump beam for pumping the optical parametric oscillators (OPO). Approximately 60 mW is passed through a mode cleaning cavity (MCC) to filter the spatial mode and is then used to make local oscillators (LO) for homodyne measurements. The rest of the beam is used for controlling optical systems such as cavity locking and phase locking of the interferometer.

The SHG used in this experiment consists of a bow-tie cavity with two spherical mirrors of curvature 50 mm with cavity length of 500 mm containing a 10 mm length KNbO\textsubscript{3} inside the cavity as the non-linear medium. With 1 W pump power, approximately 460 mW of 430 nm pump beam is generated. To ensure the spatial mode matching between the mode of the pump beam and that of the OPO, a reference cavity (not shown in Fig.~\ref{fig:expsetup}) is placed between the SHG and the OPOs.

The four OPOs used in this experiment are triangle-shaped cavities with cavity length of 48 mm and a output coupler whose transmittivity is 14 \%. A 10 mm long PPKTP crystal is placed inside the cavity\cite{Serikawa2016}. The intracavity losses for each OPO are about 0.2 to 0.3 \% (which is equivalent to 1.4 to 2.1 \% external loss). Because there are four OPOs, to ensure sufficient squeezing level with the limited pump power, we built rectangular-shaped build-up cavities wit a cavity length of 1100 mm around the OPOs to amplify the 430 nm pump beam amplitude. 

The interferometer for state generation consists of five $50:50$ beamsplitters (BS-1 to BS-5) and two optical delay lines where the length of the long delay line is an integer times the length of the short delay line. We used a free space delay line for the short delay line and fiber for the long delay line. Since the size of the wave packet employed in this experiment is 40 ns, the length of the free space delay line is 12 m and is built by using four two-inch 0-degree highly reflective mirrors. Because we are delaying five wave packets, the length of the long delay line must be equivalent to $5\times40=200\textrm{\,ns}$ delay. We use optical fiber (FFC-2PS-APC-40M-SM85-PS-H90D, Fujikura) of length 40 m for this delay line. To obtain high transmittivity through the fiber, the ends of the fiber are anti-reflective coated at 860 nm and a fiber alignment system\cite{Yokoyama2013} is employed. With this, we achieve transmittivity of about 90 \%. The interferometric visibility for the beamsplitters for state generation are all around 96-99 \%.

To control the phase of the interferometer, we inject phase reference beams into each OPO. In this experiment, the frequencies of each probe beam are shifted differently. Due to parametric amplification, the loci of the complex electric field amplitude of the probe beam in rotating reference frame of fundamental frequency becomes elliptical instead of the usual circular. The major axis of this ellipse is equivalent to the phase of the anti-squeezing quadrature and the minor axis is equivalent to the phase of the squeezing quadrature. Thus, by locking the relative orientation of the ellipses in the complex planes, we can lock the relative phases between all the beams. To shift the frequencies of each probe beam, we used acousto-optic modulators (AOM; AOMO3080-125, Crystal Technology). The RF signals necessary for driving each AOM and for demodulations must be synchronized. We used four four-channel direct digital synthesis function generators (DDS; AD9959, Analog Devices) to generate synchronized 16-channel RF signals. Synchronized clock signals of 125 MHz for DDSs are generated and distributed using homemade circuits. The demodulation frequencies for locking the relative phase between beams and the places in the setup where feedback controls are required for controlling relative phases are shown in Fig.~\ref{fig:expsetup}. There are 13 places where relative phases needed to be locked. For the feedback control of the cavity lengths for SHG, MCC, four OPOs and four building-up cavities, we use the tilt locking method (Cavity controls are omitted from Fig.~\ref{fig:expsetup}). 

The generated two-dimensional cluster state is verified by balanced-homodyne measurement. The homodyne measurement consists of a 50:50 beamsplitter and a homemade homodyne detector. This homemade homodyne detector has a bandwidth of around 100 MHz\cite{Serikawa2016}. The LO power is about 9.5 mW at each homodyne measurement, and the interferometric visibilities at the homodyne measurements are $>97$ \% . In this experiment, the probe beams are always on, thus we  electrically remove the signals due to the probe beams by using homemade 5th-order high-pass filters with cutoff frequency of 4 MHz. A low-pass filter with cutoff frequency of 100 MHz (Minicircuit, BLP-100+) is also put after the homodyne detectors for anti-aliasing.

\begin{figure}[htb]
\centering
\includegraphics[width=\textwidth]{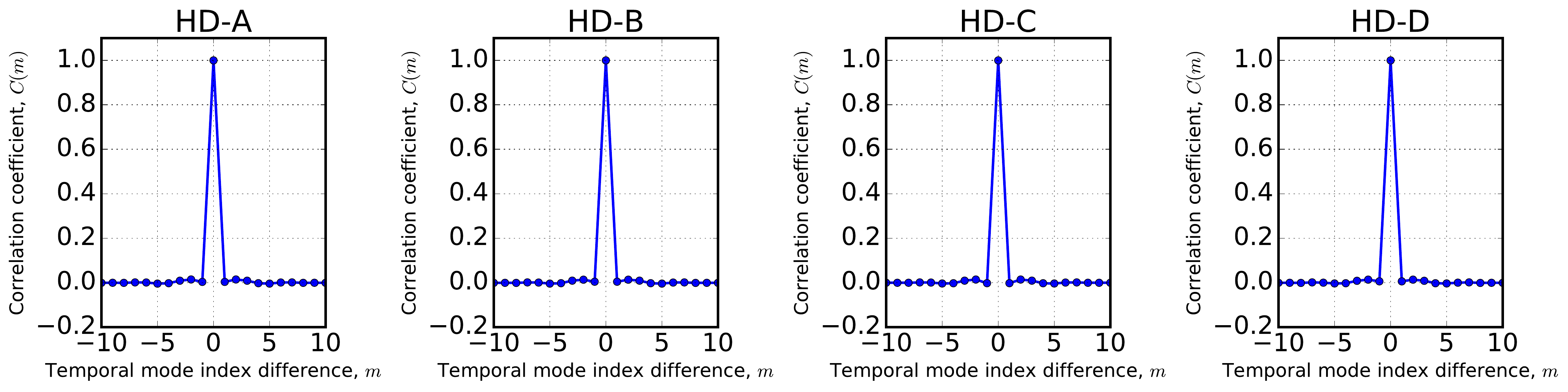}
\caption{Correlation coefficient $C(m)$ between adjacent mode function for each homodyne detector.}
\label{fig:corr}
\end{figure}

\setcitestyle{numbers,open={},close={},citesep={,}}The electrical signals from the homodyne detectors are recorded using an oscilloscope (Tektronix, DP07054). The sampling rate is set to 1 GHz and each frame has a length of 250 $\mu$s. We use 12,000 frames to calculate the statistical quantities such as variances. 40 data points are used in the calculation of quadrature values of each temporal mode. The shape of the wave packet used in this experiment is the same shape as in Ref. \cite{Yoshikawa2016}, which is designed to minimize the correlation between adjacent modes when the effect of electrical filters used in experiment and finite bandwidth of homodyne detector are taken into an account. The shape of the wave packet $f_{k}(t)$ at each temporal index $k$ is given by\setcitestyle{super,open={},close={},citesep={,}}
\begin{align}
f_{k}(t) \propto 
\begin{cases}
(t-t_{0}+0.5+t_{c})\exp[-\gamma^{2}(t-t_{0}+0.5)^{2}]&\left(\vert t-t_{0}\vert\leq\frac{\Delta t}{2}\right),\\
0 & \left(\vert t-t_{0}\vert>\frac{\Delta t}{2}\right),
\end{cases}
\end{align}
where the wave packet width is $\Delta t = 40$ ns, $t_{0}=\frac{\Delta t}{2}+(k-1)\Delta t$, $t_{c}=0$, and  $\gamma=2\pi\times 10.5$ MHz is optimized to maximize the squeezing level of the wave packet, while keeping the adjacent wave packets independent of each other. The overlap between adjacent wave packets is equal to square of the correlation coefficient $C(m)$, calculated using the formula
\begin{align}
C(m)=\frac{\langle\hat{x}_{k}^{\textrm{vac}}\hat{x}_{k+m}^{\textrm{vac}}\rangle}{\langle\hat{x}_{k}^{\textrm{vac}}\hat{x}_{k}^{\textrm{vac}}\rangle},
\end{align}
where $\hat{x}_{k}^{\textrm{vac}}$ is the quadrature operator of vacuum in the wave packet with temporal mode index $k$  \cite{Yoshikawa2016}. Figure \ref{fig:corr} shows the correlation coefficient $C(m)$ which are calculated by using shot noise measured with each homodyne detector. We can observe that the overlap between different temporal modes is negligible. This confirms that each temporal wave packet can be considered as an independent and orthogonal quantum mode that satisfies bosonic commutation relation $[\hat{a}_{k},\hat{a}_{l}^{\dagger}]=\delta_{kl}$, where $\hat{a}_{k}$ is an annihilation operator concerning wave packet mode $k$.

\section{Theory for Generation of Two-Dimensional Cluster State}
\subsection{Nullifiers and Graph}
\begin{figure}[hbt]
\centering
\includegraphics[width=\textwidth]{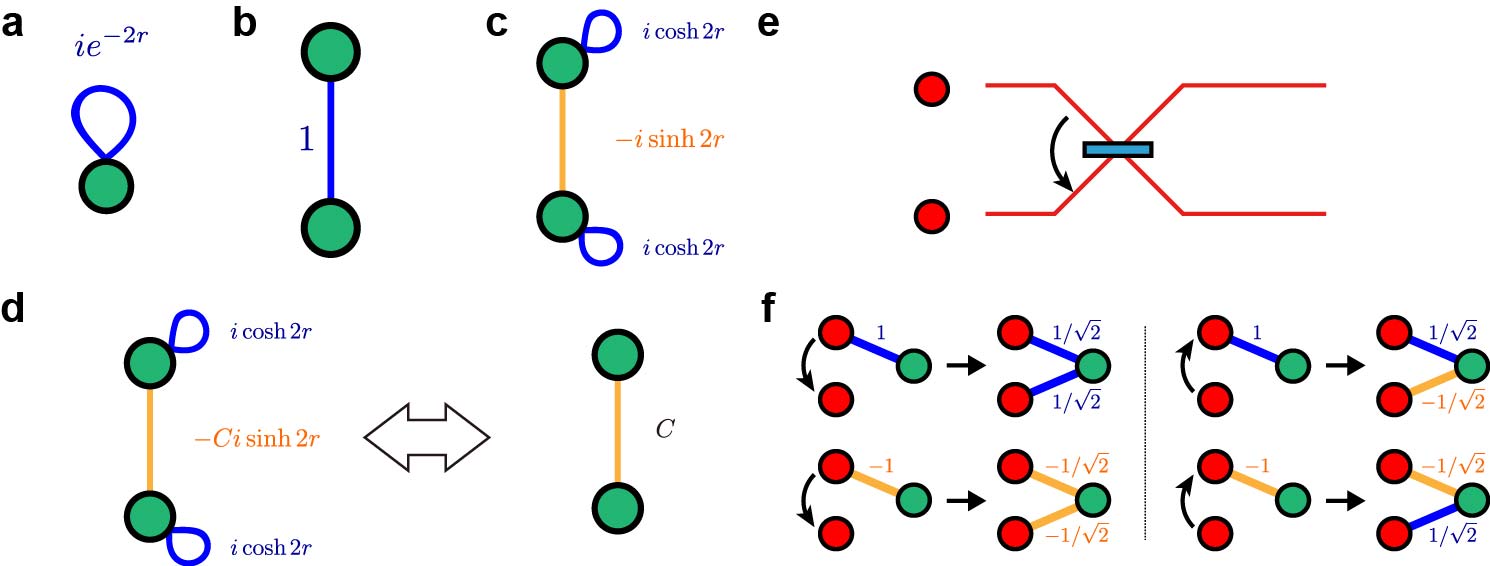}
\caption{Graph representation of Gaussian state. (a) p-squeezed vacua. (b) Ideal canonical two-mode cluster state. (c) EPR state generated from finite squeezing with squeezing parameter $r$. (d) Short-hand representation of graph used in this paper. $C$ represents the value of the edge weight factor relative to the EPR state. The color of the edge shows the sign of edge weight. (e) Beam splitter operation. (f) Graph transformation of 50:50 beam splitter for the state in this experiment.}
\label{fig:graphrule}
\end{figure}
Here, we will derive nullifiers and show the characteristics of the graph of the two-dimensional cluster state generated in this experiment. In continuous-variable quantum computation, pure Gaussian states, including cluster states, can be represented using graphs. In the graph representation, each node of the graph corresponds to a mode in the state. The graph has one-to-one correspondence with the covariance matrix of the state, which uniquely defines the Gaussian state up to local displacements\cite{Menicucci2011}. Figure \ref{fig:graphrule} (a-c) shows graph representations of some basic quantum states. In particular, when we represent the cluster state with its graph, we will use the notational shorthand shown in Fig.~\ref{fig:graphrule} (d) where self-loops are omitted and the weight edge $G=\pm Ci\sinh2r$ will be represented by only $C$ and the color of the edge corresponds to $\pm$ of the edge \setcitestyle{numbers,open={},close={},citesep={,}}(Ref. \cite{Menicucci2011a}). In the derivation of the state, only 50:50 beam splitter transformations are required and thus the graph can be simplified to Fig.~\ref{fig:graphrule} (f). The arrow represents the phase relationship of the beam splitter transformation. For more details on the graph representation see Ref. \cite{Menicucci2011}.\setcitestyle{super,open={},close={},citesep={,}}

\begin{figure}[htbp]
\centering
\includegraphics[width=0.9\textwidth]{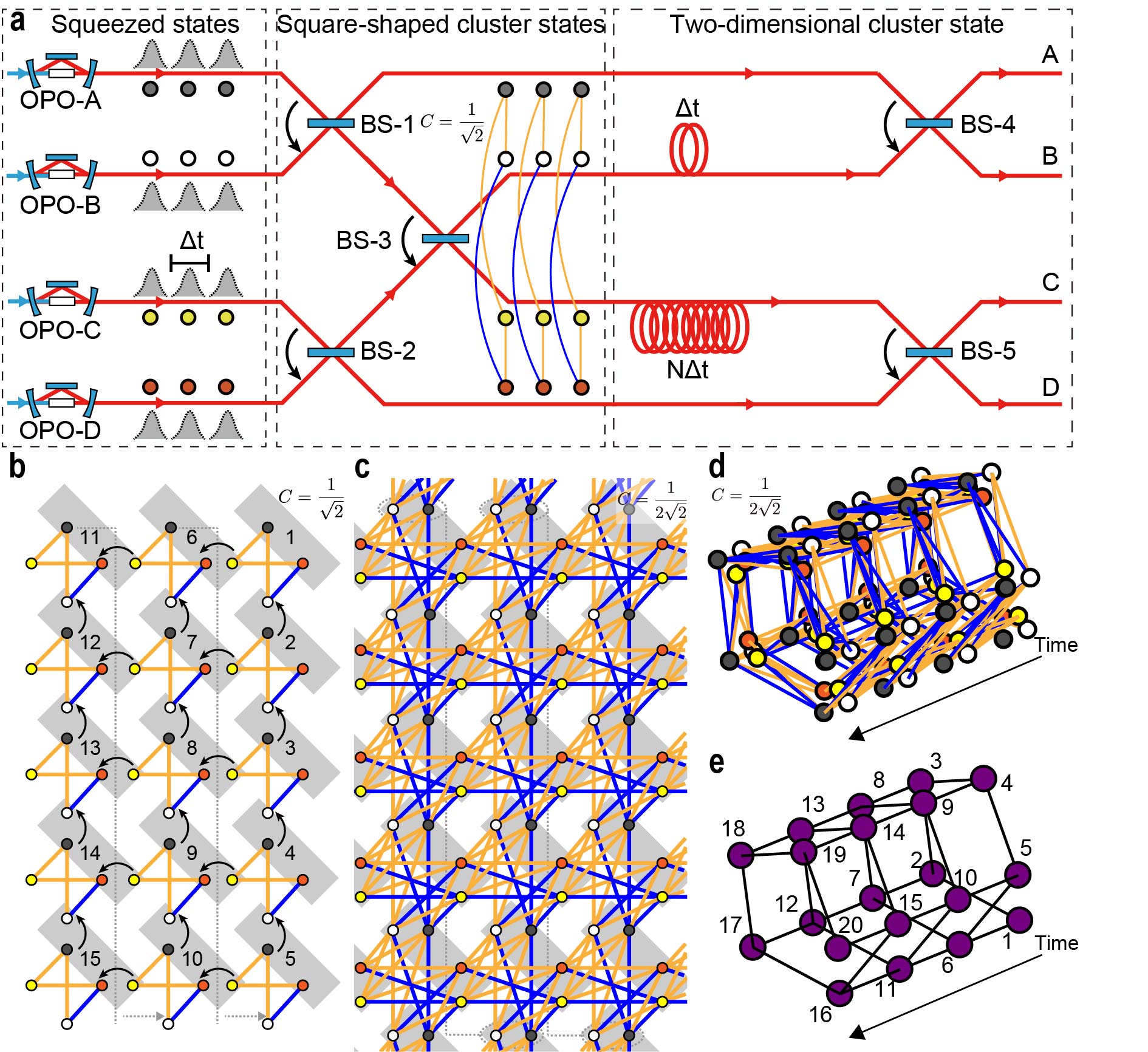}
\caption{Graph of two-dimensional cluster state. (a) System for state generation showing square-shaped cluster states after first three beam splitters. (b) After square-shaped cluster state are generated, they are connected into two-dimensional cluster state using two optical delay lines and two beam splitters. (c) Resulting graph of the state. (d) Resulting graph of the state showing actual cylindrical boundary condition. (e) Corresponding temporal indices for the graph. The purple nodes represent the logical node of the resulting state. For (d), (e) we show 80 light modes (20 temporal indices).}
\label{fig:graph}
\end{figure}

First, let us perform the derivation in the ideal case of infinite squeezing. If the state from the OPO-A and OPO-C are $\ket{p=0}$ and the state from OPO-B and OPO-D are $\ket{x=0}$, the initial nullifiers for each wave packet with temporal index $k$ become
\begin{align}
\{\hat{p}_{k}^{A},\hat{x}_{k}^{B},\hat{p}_{k}^{C},\hat{x}_{k}^{D}\}.
\end{align}
After passing through the first three beamsplitters, BS-1 to BS-3, the nullifiers become
\begin{align}
\left\{\hat{p}_{k}^{A}+\frac{\hat{p}_{k}^{B}+\hat{p}_{k}^{C}}{\sqrt{2}},\hat{x}_{k}^{A}-\frac{\hat{x}_{k}^{B}+\hat{x}_{k}^{C}}{\sqrt{2}},\hat{p}_{k}^{D}+\frac{-\hat{p}_{k}^{B}+\hat{p}_{k}^{C}}{\sqrt{2}},\hat{x}_{k}^{D}-\frac{-\hat{x}_{k}^{B}+\hat{x}_{k}^{C}}{\sqrt{2}}\right\}.
\end{align}
At this point, the graph of the state at each temporal index $k$ is a square where the wave packet of the four spatial modes are entangled as shown in Fig.~\ref{fig:graph} (a). Finally, the two optical delay lines and beamsplitters BS-4 and BS-5 entangle the state into a two-dimensional cluster state with nullifiers
\begin{subequations}
\begin{align}
\hat{\delta}_{k}^{(p,1)}&=\hat{p}_{k}^{A}+\hat{p}_{k}^{B}+\frac{1}{\sqrt{2}}\left(-\hat{p}_{k+1}^{A}+\hat{p}_{k+1}^{B}+\hat{p}_{k+N}^{C}+\hat{p}_{k+N}^{D}\right),\\
\hat{\delta}_{k}^{(x,1)}&=\hat{x}_{k}^{A}+\hat{x}_{k}^{B}-\frac{1}{\sqrt{2}}\left(-\hat{x}_{k+1}^{A}+\hat{x}_{k+1}^{B}+\hat{x}_{k+N}^{C}+\hat{x}_{k+N}^{D}\right),\\
\hat{\delta}_{k}^{(p,2)}&=\hat{p}_{k}^{C}-\hat{p}_{k}^{D}+\frac{1}{\sqrt{2}}\left(-\hat{p}_{k+1}^{A}+\hat{p}_{k+1}^{B}-\hat{p}_{k+N}^{C}-\hat{p}_{k+N}^{D}\right),\\
\hat{\delta}_{k}^{(x,2)}&=\hat{x}_{k}^{C}-\hat{x}_{k}^{D}-\frac{1}{\sqrt{2}}\left(-\hat{x}_{k+1}^{A}+\hat{x}_{k+1}^{B}-\hat{x}_{k+N}^{C}-\hat{x}_{k+N}^{D}\right).
\end{align}
\end{subequations}
Note that the nullifiers of our state have a different form from the nullifiers in the original CV cluster state proposal, which are of the form $\hat{\delta}=\hat{p}-\sum g_{i}\hat{x}_{i}$. However, the nullifiers of our state can also be transformed into $\hat{p}-\sum g_{i}\hat{x}_{i}$ form by applying local Fourier transforms (a.k.a., $\frac{\pi}{2}$ phase rotations) on half of the modes of the cluster state. This relations means the computational capability of our state is no different from that of the original CV cluster state and only difference is a change in the measurement basis. The graph of the resulting cluster state is shown in Fig.~\ref{fig:graph} (c) for the case where $N=5$. The absolute value of the edge weight of this graph is $C=\frac{1}{2\sqrt{2}}$ times of that of EPR state. Note that in case of finite squeezing, in addition to edges between nodes, there is also self loop whose weights are exactly the same as in the case of the EPR state. Because applying Fourier transforms on half of the modes does not change the structure of the graph besides the edge weights, even if we recast the nullifiers into $\hat{p}-\sum g_{i}\hat{x}_{i}$ form, the structure of the graph will be the same but with different edge weights. Also note that the boundary condition of this graph is an infinitely long helix, where the number of the nodes on each turn of the helix corresponds to $N$, which is the number of inputs that can be used in this cluster state. The infinite length means this system can generate cluster states with an unlimited number of modes, which corresponds to unlimited number of operations in either the limit of infinite squeezing or if proper error correction is used.
In the case of finite squeezing, the values of the above nullifiers become non-zero and can be expressed in the Heisenberg picture as
\begin{subequations}
\begin{align}
\hat{\delta}_{k}^{(p,1)}&=2\hat{p}_{k}^{A,\textrm{vac}}e^{-r_{A}},\\
\hat{\delta}_{k}^{(x,1)}&=2\hat{x}_{k}^{B,\textrm{vac}}e^{-r_{B}},\\
\hat{\delta}_{k}^{(p,2)}&=2\hat{p}_{k}^{C,\textrm{vac}}e^{-r_{C}},\\
\hat{\delta}_{k}^{(x,2)}&=2\hat{x}_{k}^{D,\textrm{vac}}e^{-r_{D}},
\end{align}
\end{subequations}
where $r_{j}$ is squeezing parameter of squeezed light from OPO $j$ and $\hat{x}_{k}^{\textrm{vac}},\hat{p}_{k}^{\textrm{vac}}$ are quadratures of vacuum state. This means that in the ideal case without optical loss and experimental imperfection, the value of the nullifiers of the cluster state generated with finite squeezing resources corresponds to the original squeezing from each OPO.
\subsection{Derivation of inseparability criteria}

\setcitestyle{numbers,open={},close={},citesep={,}}We will derive criteria and a threshold of variance for the the nullifiers to establish quantum inseparability. For this cluster state, the resulting criteria is that if the variance of each nullifier is observed to be below $-4.5$ dB comparing to shot noise, then the inseparability of the state can be established and we can verify the state.  In the one-dimensional case, the threshold for inseparability is $-3$ dB with respect to shot noise. The increment in the threshold is due to the higher complexity and dimensions of two-dimensional cluster state. As in Ref. \cite{Yokoyama2013}, we can derive inseparability criteria using van Loock-Furusawa criteria\setcitestyle{super,open={},close={},citesep={,}}\cite{Loock2003}. This criteria is as follows. First, we let the set of the modes of the state be $M$. Next, we separate this set into two set $A$ and $B$ with $A\cup B=M$ and $A\cap B=\phi$. When we consider the linear combination of quadratures $\hat{X}=\sum_{j}h_{j}\hat{x}_{j}$, $\hat{P}=\sum_{k}g_{k}\hat{p}_{k}$, if the sum of the variances satisfy
\begin{align}
\mean{(\Delta\hat{X})^{2}}+\mean{(\Delta\hat{P})^{2}}<\hbar\left(\sum_{a\in A}\abs{g_{a}h_{a}}+\sum_{b\in B}\abs{g_{b}h_{b}}\right),\label{eq:vanloock}
\end{align}
then the state $\hat{\rho}$ cannot be represent as $\hat{\rho}=\sum_{i}p_{i}\hat{\rho}_{i}^{A}\otimes\hat{\rho}_{i}^{B}$, where $\hat{\rho}_{i}^{A},\hat{\rho}_{i}^{B}$ are the state in the subspace of $A$ and $B$, respectively.

\begin{figure}[htb]
\centering
\includegraphics[width=\textwidth]{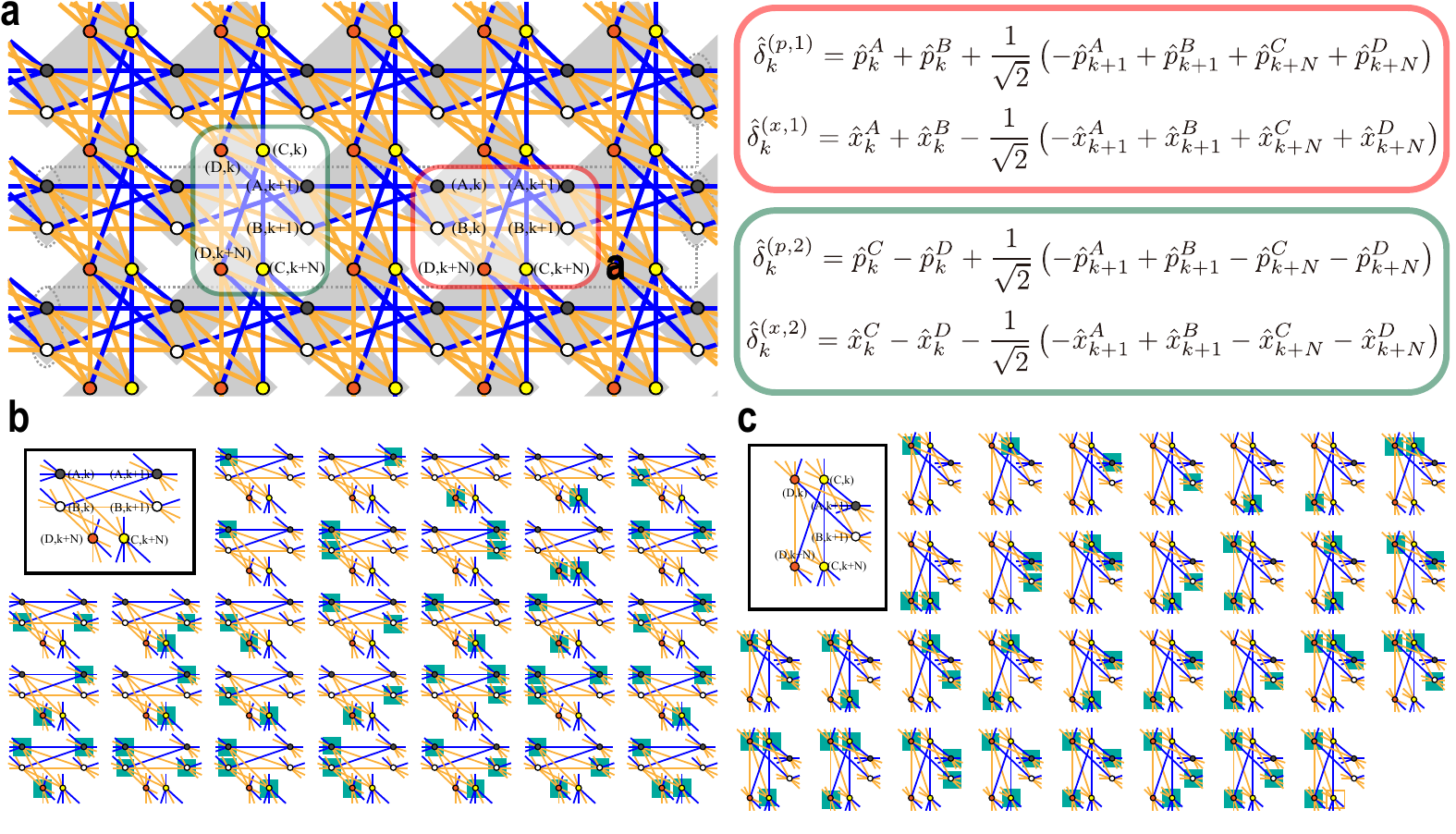}
\caption{Inseparability of two-dimensional cluster state with nullifiers. (a) Two types of nullifiers connecting six temporal modes in two-direction. (b) 31 possible bipartitions of six temporal modes connected by $\hat{\delta}_{k}^{(x,1)}$ and $\hat{\delta}_{k}^{(p,1)}$. (c) 31 possible bipartitions of six temporal modes connected by $\hat{\delta}_{k}^{(x,2)}$ and $\hat{\delta}_{k}^{(p,2)}$. For (b) and (c) the nodes with green backgrounds are the node in the same bipartition.}
\label{fig:inseparability}
\end{figure}

Therefore, if we use the van Loock-Furusawa criterion, where the linear combination of quadratures here is given by the nullifiers of the cluster state, then there exists a finite squeezing threshold for inseparability criteria. Furthermore, we can show inseparability by  experimentally measuring nullifiers and checking whether the variances of nullifiers are below that threshold or not. In Ref. \setcitestyle{numbers,open={},close={},citesep={,}}\cite{Yokoyama2013}, by doing this repeatedly for each temporal index $k$, the authors show that the generated state is indeed inseparable and is connected in one-dimension. In the two-dimension case, however, this has to be done in two directions. As it is shown in Fig.~\ref{fig:inseparability}, $\hat{\delta}_{k}^{(p,1)}$ and $\hat{\delta}_{k}^{(x,1)}$ connect the nodes in the circumference direction while $\hat{\delta}_{k}^{(p,2)}$ and $\hat{\delta}_{k}^{(x,2)}$ connect the nodes in the direction along the infinite length of the helix. To show inseparability between nodes concerning nullifiers of each $k$, 62 types of bipartition must be negated. In Ref. \cite{Yokoyama2013}, there were only 7 possible bipartitions, thus it was feasible to perform the calculation directly. In our case, we perform the calculation systematically. In this paper, we will show the calculations for modes concerning $\hat{\delta}_{k}^{(p,1)}$ and $\hat{\delta}_{k}^{(x,1)}$ only. The other half can be derived in a similar manner.

Let us use notation $(i,k)$ to represent the quadrature of the wave packet at spatial mode $i$ with temporal mode index $k$. Bipartition of six modes can be separated into three types: one mode and five modes, two modes and four modes, three modes and three modes. Let us consider the condition for each case separately. To derive the threshold more smoothly, let us make following simplification: for any particular bipartition and linear combination, if Eq. \eqref{eq:vanloock} is of the form
\begin{align}
\mean{(\Delta\hat{X})^{2}}+\mean{(\Delta\hat{P})^{2}}<\hbar A,
\end{align}
then the sufficient condition to satisfy the above inequality is
\begin{align}
\mean{(\Delta\hat{X})^{2}}<A\frac{\hbar}{2}\quad\textrm{and}\quad\mean{(\Delta\hat{P})^{2}}<A\frac{\hbar}{2}.
\end{align}
In the actual experiment, we normalize the nullifiers with respect to vacuum. Thus, if we supposed that the variance for the linear combination of the quadratures for the vacuum states is $\mean{(\Delta\hat{X})^{2}}_{0}$, $\mean{(\Delta\hat{P})^{2}}_{0}$, then the threshold with respect to shot noise in dB is
\begin{align}
\textrm{(Threshold in \# of dB)}=10\log_{10}\left(\frac{A\hbar}{2\mean{(\Delta\hat{X})^{2}}_{0}}\right).
\end{align}
Note that the similar relation holds for $\hat{P}$. In our particular form of nullifiers, $\mean{(\Delta\hat{X})^{2}}_{0}$ equals to four units of shot noise, i.e. $4\times\frac{\hbar}{2}$, thus the above is simplified to
\begin{align}
\textrm{(Threshold in \# of dB)}=10\log_{10}\left(\frac{A}{4}\right).
\end{align}
\subsubsection{One modes and five modes}

\begin{enumerate}
\item for the case where one mode is $(A,k)$ or $(B,k)$, we get
\begin{align}
\mean{(\Delta\hat{\delta}_{k}^{(x,1)})^{2}}+\mean{(\Delta\hat{\delta}_{k}^{(p,1)})^{2}}<2\hbar,
\end{align}
which corresponds to 3 dB of squeezing for each nullifier.
\item for the case where one mode is $(A,k+1)$ or $(B,k+1)$, we get
\begin{align}
\mean{(\Delta\hat{\delta}_{k}^{(x,1)})^{2}}+\mean{(\Delta\hat{\delta}_{k+1}^{(p,1)})^{2}}<\sqrt{2}\hbar,
\end{align}
which corresponds to 4.5 dB of squeezing for each nullifier.
\item for the case where one mode is $(C,k+N)$ or $(D,k+N)$, we get
\begin{align}
\mean{(\Delta\hat{\delta}_{k}^{(x,1)})^{2}}+\mean{(\Delta\hat{\delta}_{k}^{(p,2)})^{2}}<\sqrt{2}\hbar,
\end{align}
which corresponds to 4.5 dB of squeezing for each nullifier.
\end{enumerate}
\subsubsection{Two modes and four modes}
\begin{enumerate}
\item for the case where the two modes are $(A,k)$ and $(B,k)$, we get
\begin{align}
\mean{(\Delta\hat{\delta}_{k}^{(x,1)})^{2}}+\mean{(\Delta\hat{\delta}_{k}^{(p,1)})^{2}}<4\hbar,
\end{align}
which corresponds to 0 dB of squeezing for each nullifier.
\item for the case where the two modes are from $(A,k+1)$, $(B,k+1)$, $(C,k+N)$, and $(D,k+N)$, we get
\begin{align}
\mean{(\Delta\hat{\delta}_{k}^{(x,1)})^{2}}+\mean{(\Delta\hat{\delta}_{k}^{(p,1)})^{2}}<2\hbar,
\end{align}
which corresponds to 3 dB of squeezing for each nullifier.
\item for the rest of the cases not belonged to the above, we get
\begin{align}
\mean{(\Delta\hat{\delta}_{k-1}^{(x,1)})^{2}}+\mean{(\Delta\hat{\delta}_{k}^{(p,1)})^{2}}<\sqrt{2}\hbar,
\end{align}
which corresponds to 4.5 dB of squeezing for each nullifier.
\end{enumerate}
\subsubsection{Three modes and three modes}
\begin{enumerate}
\item for the case where two of the modes are $(A,k)$ and $(B,k)$ and the other mode are one from $(A,k+1)$, $(B,k+1)$, $(C,k+N)$, and $(D,k+N)$, we get 
\begin{align}
\mean{(\Delta\hat{\delta}_{k}^{(x,1)})^{2}}+\mean{(\Delta\hat{\delta}_{k}^{(p,1)})^{2}}<3\hbar,
\end{align}
which corresponds to 1.2 dB of squeezing for each nullifier.
\item for the case where three modes are from $(A,k+1)$, $(B,k+1)$, $(C,k+N)$, and $(D,k+N)$, we get 
\begin{align}
\mean{(\Delta\hat{\delta}_{k}^{(x,1)})^{2}}+\mean{(\Delta\hat{\delta}_{k}^{(p,1)})^{2}}<3\hbar,
\end{align}
which corresponds to 1.2 dB of squeezing for each nullifier.
\item for the case not belong to the above, we get
\begin{align}
\mean{(\Delta\hat{\delta}_{k-1}^{(x,1)})^{2}}+\mean{(\Delta\hat{\delta}_{k}^{(p,1)})^{2}}<\sqrt{2}\hbar,
\end{align}
which corresponds to 4.5 dB of squeezing for each nullifier.
\end{enumerate}

Therefore, we can conclude that 4.5 dB of squeezing for each nullifier is a sufficient conditions for inseparability. Note that since this condition is merely sufficient, there is a possibility that there might exist a condition with a lower squeezing threshold. However, our motivation for showing that the criteria is 4.5 dB for this state is that in the case of a Gaussian cluster states with nullifiers of the form $\hat{\mathbf{p}}-G\hat{\mathbf{x}}$ whose graph is bipartite and self-inverse, the necessary squeezing level in dB for showing inseparability is $-10\log_{10}g$ where $g$ is the weight edge. This can be derived by further extending van Loock-Furusawa criteria and using the relations in Ref. \cite{Ukai2015}. First, for nullifiers of the form $\delta_{i}=\hat{p}_{i}-\sum_{k}G_{ik}\hat{x}_{k}$, the variance of the nullifiers when the squeezing level is finite becomes
\begin{align}
\mean{(\Delta\hat{\delta}_{k})^{2}}=\frac{\hbar}{2}e^{-2r}[1+g^{2}M_{N(j)}],
\end{align}
where $g$ is the weight edge which is assumed to be the same for all edges up to plus-minus signs. By using the bipartite and self-inverse property of $G$, we can show that the separability between $k$ and $l$ can be shown if
\begin{align}
\mean{(\Delta\hat{\delta}_{k})^{2}}+\mean{(\Delta\hat{\delta}_{l})^{2}}<2\hbar G_{jl}.
\end{align}
The necessary squeezing level in dB determined by the above relation is $-10\log_{10}g$. Since our state is equivalent to a Gaussian cluster state up to local phase shifts with $g=\frac{1}{2\sqrt{2}}$, we expect that even in the worst case, 4.5 dB of squeezing is necessary for showing inseparability and we have shown explicitly in this section that this is indeed the case.

\section{Measurement of the length of two optical delay line}
One of the main differences between the generation systems for the two-dimensional cluster state and the one-dimensional cluster state is that the former has two optical delay lines, where the length of the long one is an integer times of the short one. In the one-dimensional case, the optical delay line does not need to be built with precise length; we can pick the wave packet whose width matches the length of the optical delay line built. For the two-dimensional cluster state, this is not the case due to the restriction aforementioned. Thus, we need to build a system where both optical delay lines can be measured and tuned precisely. This is the main reason we prefer the current generation system over Ref.~\cite{Alexander2018} where one of the optical delay line cannot be measured precisely with our interferometric technique. The ability to measure and tune optical delay line is important especially when we consider using shorter wave packet widths and higher squeezing levels, which would require higher precision in matching the length of the optical delay line.

\begin{figure}[htb]
\centering
\includegraphics[width=\textwidth]{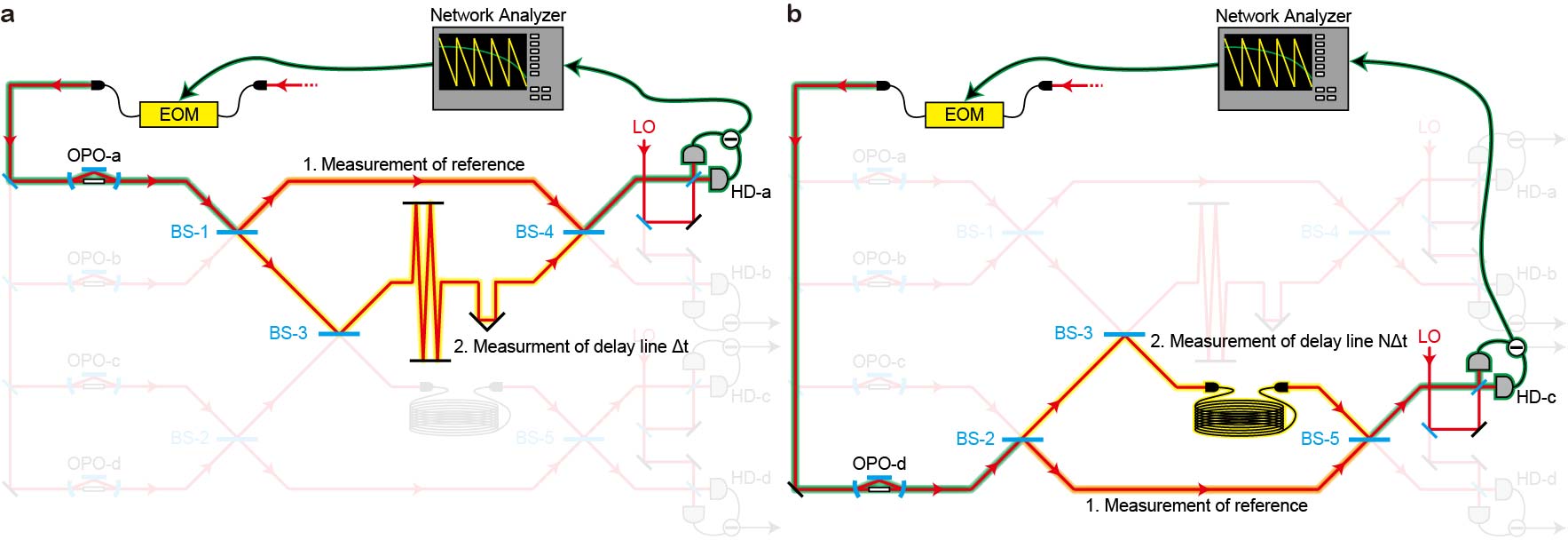}
\caption{Measurement of the optical delay lines. (a) Short delay line. (b) Long delay line. The figures show the only the part of the setup that are involved in this measurement. The measurements is done using Electro-optic modulator and measure the phase shift of each frequency when the modulated signals go through the reference path (Network analyzer$\rightarrow$green path$\rightarrow$orange path$\rightarrow$green path$\rightarrow$Network analyzer) and optical delay path (Network analyzer$\rightarrow$green path$\rightarrow$yellow path$\rightarrow$green path$\rightarrow$Network analyzer). Then, by comparing the phase difference between the two measurement results, we can find the length of the optical delay line (difference between yellow path and orange path).}
\label{fig:phasemeasurementsetup}
\end{figure}

\begin{figure}[hbt]
\centering
\includegraphics[width=\textwidth]{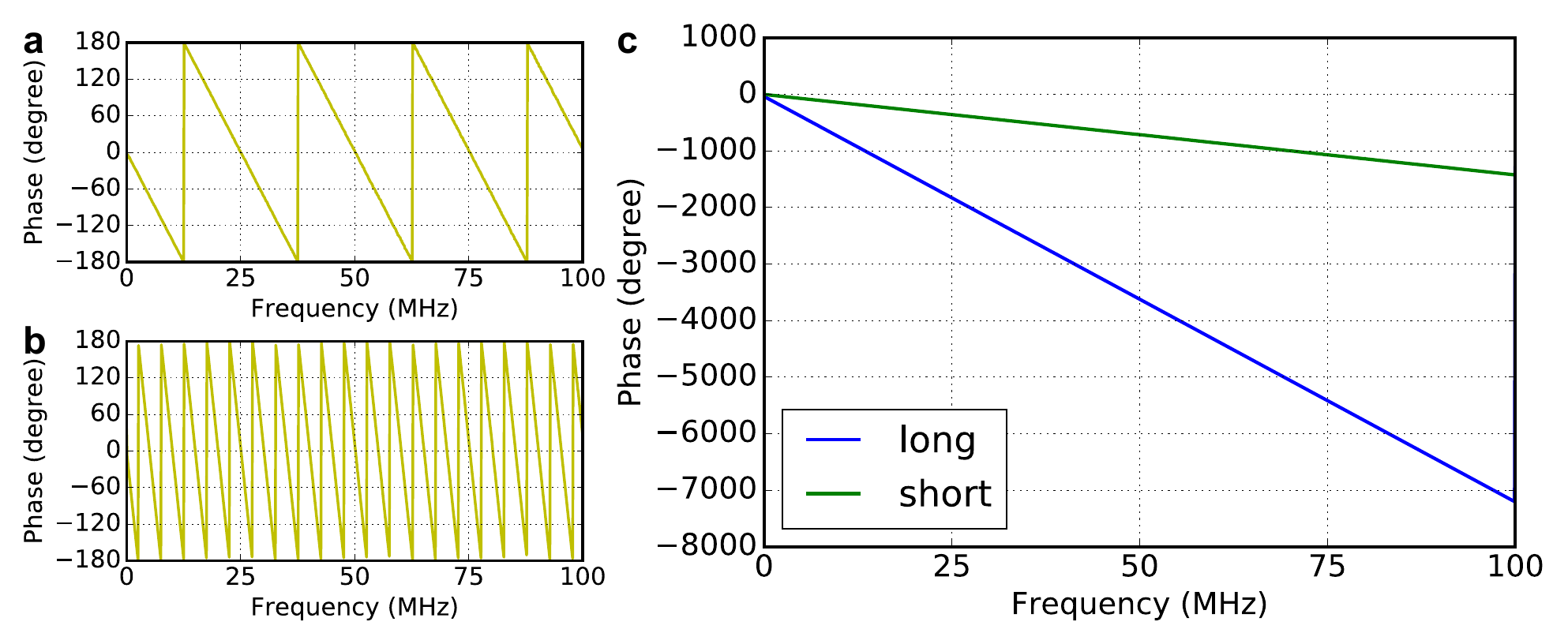}
\caption{Measurement results of the length of each optical delay line via phase difference measurements. The length of the optical delay line is determined by fitting the slope of the phase-frequency dependence. (a) Phase difference for short optical delay line. (b) Phase difference for long optical delay line. (c) Plots of phase difference for both delay lines.}
\label{fig:phasemeasurement}
\end{figure}

To measure the length of each optical delay line, we perform the measurements with the scheme shown in Fig.~\ref{fig:phasemeasurementsetup}. By measuring the frequency dependence of the phase difference between two paths, we can measure the length of the optical delay line, and then, we use the optical stage to calibrate the length. The resulting phase measurement is shown in Fig.~\ref{fig:phasemeasurement} and the length measurement results are 11.85 m (39.6 ns) and 59.75 m (199.8 ns). The ratio between them is $0.99:5$.


\section{Effects of optical losses and length mismatch on nullifiers}
In this section, we will explain how optical losses and length mismatch of the two optical delay lines affect nullifiers. As opposed to the one-dimensional case, where the experimental setup is a simple asymmetric Mach-Zehnder interferometer, the setup for two-dimensional case is a more complex interferometer with two optical delay lines of which the length of the longer delay line should ideally be an integer multiple of the short one. Therefore, how the optical losses in each path affect each of the nullifiers and what happens when the length of the two delay lines deviate from an integer multiple of each other is more complicated. The latter, especially, is a problem that does not arise in the one-dimensional case.

When we consider the values of nullifiers theoretically, it is useful to use frequency domain for calculations. Let us suppose we encode the quantum state in a wave packet $f(t)$ whose Fourier transform is $\tilde{f}(\omega)$ and both of them are normalized as
\begin{align}
\int_{-\infty}^{\infty}\textrm{d}t\,\abs{f(t)}^{2}=\int_{-\infty}^{\infty}\textrm{d}\omega\,\abs{\tilde{f}(\omega)}^{2}=1.
\end{align}
Then, the value of the nullifiers when we take into account the optical losses and the length mismatch can be calculated as
\begin{subequations}
\begin{align}
\begin{split}
\mean{(\Delta\hat{\delta}_{k}^{(x,1)})^{2}}=&\eta\int\abs{\tilde{f}(\omega)}^{2}\Bigg\{\abs{\frac{1-e^{i\omega\Delta\tau_{2}}}{2}}^{2}S_{+}^{A}(\omega)+\abs{\frac{3+e^{i\omega\Delta\tau_{2}}}{2}}^{2}S_{-}^{B}(\omega)\\
&+\abs{\frac{1-e^{i\omega\Delta\tau_{2}}}{2}}^{2}\left[S_{+}^{C}(\omega)+S_{-}^{D}(\omega)\right]\Bigg\}\textrm{d}\omega+4(1-\eta),
\end{split}\\
\begin{split}
\mean{(\Delta\hat{\delta}_{k}^{(p,1)})^{2}}=&\eta\int\abs{\tilde{f}(\omega)}^{2}\Bigg\{\abs{\frac{3+e^{i\omega\Delta\tau_{2}}}{2}}^{2}S_{-}^{A}(\omega)+\abs{\frac{1-e^{i\omega\Delta\tau_{2}}}{2}}^{2}S_{+}^{B}(\omega)\\
&+\abs{\frac{1-e^{i\omega\Delta\tau_{2}}}{2}}^{2}\left[S_{-}^{C}(\omega)+S_{+}^{D}(\omega)\right]\Bigg\}\textrm{d}\omega+4(1-\eta),
\end{split}\\
\begin{split}
\mean{(\Delta\hat{\delta}_{k}^{(x,2)})^{2}}=&\eta\int\abs{\tilde{f}(\omega)}^{2}\Bigg\{\abs{\frac{1-e^{i\omega\Delta\tau_{2}}}{2}}^{2}S_{+}^{C}(\omega)+\abs{\frac{3+e^{i\omega\Delta\tau_{2}}}{2}}^{2}S_{-}^{D}(\omega)\\
&+\abs{\frac{1-e^{i\omega\Delta\tau_{2}}}{2}}^{2}\left[S_{+}^{A}(\omega)+S_{-}^{B}(\omega)\right]\Bigg\}\textrm{d}\omega+4(1-\eta),
\end{split}\\
\begin{split}
\mean{(\Delta\hat{\delta}_{k}^{(p,2)})^{2}}=&\eta\int\abs{\tilde{f}(\omega)}^{2}\Bigg\{\abs{\frac{3+e^{i\omega\Delta\tau_{2}}}{2}}^{2}S_{-}^{C}(\omega)+\abs{\frac{1-e^{i\omega\Delta\tau_{2}}}{2}}^{2}S_{+}^{D}(\omega)\\
&+\abs{\frac{1-e^{i\omega\Delta\tau_{2}}}{2}}^{2}\left[S_{-}^{A}(\omega)+S_{+}^{B}(\omega)\right]\Bigg\}\textrm{d}\omega+4(1-\eta).
\end{split}
\end{align}
\end{subequations}
Here we assumed that when the nullifiers are calculated, the width of the wave packet $\Delta t$ is picked to be the same as the time delay of the short delay lines and $\Delta \tau_{2}=\tau_{2}-N\Delta t$ with $\tau_{2}$ being the actual time delay due to long delay line. $\eta$ is the detection efficiency which is assumed to be equal for all paths. $S_{-}(\omega)$ and $S_{+}(\omega)$ are squeezing and anti-squeezing spectral functions from the OPO and are given by
\begin{align}
S_{\pm}(\omega)=\frac{\hbar}{2}\left[1\pm\frac{T}{T+L}\frac{4\xi}{(1\mp \xi)^{2}+(\omega/\omega_{0})^{2}}\right],
\end{align}
where $T$ is the transmittivity of the output coupler of the OPO, $\xi$ is the normalized pump amplitude, $\omega_{0}$ is the bandwidth of the OPO, and $L$ is the intracavity loss.

In the ideal case where $\Delta \tau_{2} = 0$ the above equations can be simplified to 
\begin{align}
\mean{(\Delta\hat{\delta}_{k}^{(x,1)})^{2}}=&4\int\abs{\tilde{f}(\omega)}^{2}\left[\eta S_{-}^{B}(\omega)+(1-\eta)\right]\textrm{d}\omega,\\
\mean{(\Delta\hat{\delta}_{k}^{(p,1)})^{2}}=&4\int\abs{\tilde{f}(\omega)}^{2}\left[\eta S_{-}^{A}(\omega)+(1-\eta)\right]\textrm{d}\omega,\\
\mean{(\Delta\hat{\delta}_{k}^{(x,2)})^{2}}=&4\int\abs{\tilde{f}(\omega)}^{2}\left[\eta S_{-}^{D}(\omega)+(1-\eta)\right]\textrm{d}\omega,\\
\mean{(\Delta\hat{\delta}_{k}^{(p,2)})^{2}}=&4\int\abs{\tilde{f}(\omega)}^{2}\left[\eta S_{-}^{C}(\omega)+(1-\eta)\right]\textrm{d}\omega,
\end{align}
which is equivalent to additional loss of $1-\eta$ to the source of the squeezed light. Thus, when there exists some additional length mismatch, the additional anti-squeezing is mixed into the nullifiers and degrades them. Also, $\Delta\tau_{2}\neq 0$ will result in degradation of orthogonality and independence of the adjacent temporal modes. While we assume the detection efficiency $\eta$ to be the same for all paths in the above derivations, the actual values vary with the path and the beam. We measure the detection efficiency for all beams and all paths, resulting in detection efficiency of about 0.75 to 0.80 depending on which path the beam goes through. The normalized pump amplitudes of the OPOs is measured with parametric amplification of coherent light which results in about $\xi=0.65$. If we assume that there is no length mismatch, for $\eta=0.75$ we predict nullifier value of $-4.96$ dB, while for $\eta=0.80$ we predict nullifier value of $-5.62$ dB. Both predictions are in good agreement with the experimental results. The actual situation is much more complex due to the fact that the detection efficiency is not uniform over all paths, which would result in additional contamination of the anti-squeezing components. Moreover, with many relative phases that need to be locked (4 relative phase of pump beams, 5 beam splitter for state generation, 4 beam splitter for homodyne measurements), the offset of the phase at each beam splitter and phase fluctuation at each beam splitter would degrade the nullifiers. However, our prediction of nullifiers is already in good agreement with the experimental results even without taking any phase related noise and length mismatch into consideration.                                                                                                                                                                                                                                                                                                                                                                                                                                                                                                                                                                                                                                                                                                                                              

\section{Power spectral from homodyne detectors}

In this section, we will discuss the power spectrum from each homodyne detector when the two-dimensional cluster state is measured. Despite the fact that the quantum state is encoded in temporal wave packets in this experiment, a frequency domain analysis of the state gives many theoretical and experimental insights into the generated state. In particular, since our system is a complex interferometer composed of four light sources and nine beam splitters (five for state generation and four for state verification), we need a simple method for checking whether the relative phases between all beams are locked correctly or not. One might argue that this could be done by checking nullifiers, however, that is not the case. In experiment, if we perform the measurements of the nullifiers and the values are bad, the values of the nullifiers do not give us any information on which beam splitter's phase locking is incorrect. Moreover, since quadratures $\hat{x}$ and $\hat{p}$ are defined by the relative phase of the probe beams, even if the relative phases at each beam splitter are incorrect and an incorrect state is generated, there is a possibility that we will observe squeezed nullifiers based on a measurement in the wrong basis. Thus, power spectral analysis provides us with simple but powerful way of experimentally checking state generation. If we calculate the quadrature signal detected at each homodyne detector and perform Fourier transformation of the signal, the power spectral of the signal in the frequency domain can be expressed as
\begin{subequations}
\begin{align}
\begin{split}
\mean{\hat{X}_{A}^{2}(\omega)}&=\left[\frac{3}{8}-\frac{1}{2\sqrt{2}}\cos(\omega\tau_{1})\right]\eta S_{+}^{A}(\omega)+\left[\frac{3}{8}+\frac{1}{2\sqrt{2}}\cos(\omega\tau_{1})\right]\eta S_{-}^{B}(\omega)\\
&+\frac{\eta}{8}\left[S_{+}^{C}(\omega)+S_{-}^{D}(\omega)\right]+1-\eta,
\end{split}\\
\begin{split}
\mean{\hat{P}_{A}^{2}(\omega)}&=\left[\frac{3}{8}-\frac{1}{2\sqrt{2}}\cos(\omega\tau_{1})\right]\eta S_{-}^{A}(\omega)+\left[\frac{3}{8}+\frac{1}{2\sqrt{2}}\cos(\omega\tau_{1})\right]\eta S_{+}^{B}(\omega)\\
&+\frac{\eta}{8}\left[S_{-}^{C}(\omega)+S_{+}^{D}(\omega)\right]+1-\eta,
\end{split}\\
\begin{split}
\mean{\hat{X}_{B}^{2}(\omega)}&=\left[\frac{3}{8}+\frac{1}{2\sqrt{2}}\cos(\omega\tau_{1})\right]\eta S_{+}^{A}(\omega)+\left[\frac{3}{8}-\frac{1}{2\sqrt{2}}\cos(\omega\tau_{1})\right]\eta S_{-}^{B}(\omega)\\
&+\frac{\eta}{8}\left[S_{+}^{C}(\omega)+S_{-}^{D}(\omega)\right]+1-\eta,
\end{split}\\
\begin{split}
\mean{\hat{P}_{B}^{2}(\omega)}&=\left[\frac{3}{8}+\frac{1}{2\sqrt{2}}\cos(\omega\tau_{1})\right]\eta S_{-}^{A}(\omega)+\left[\frac{3}{8}-\frac{1}{2\sqrt{2}}\cos(\omega\tau_{1})\right]\eta S_{+}^{B}(\omega)\\
&+\frac{\eta}{8}\left[S_{-}^{C}(\omega)+S_{+}^{D}(\omega)\right]+1-\eta,
\end{split}\\
\begin{split}
\mean{\hat{X}_{C}^{2}(\omega)}&=\left[\frac{3}{8}-\frac{1}{2\sqrt{2}}\cos(\omega\tau_{2})\right]\eta S_{+}^{C}(\omega)+\left[\frac{3}{8}+\frac{1}{2\sqrt{2}}\cos(\omega\tau_{2})\right]\eta S_{-}^{D}(\omega)\\
&+\frac{\eta}{8}\left[S_{+}^{A}(\omega)+S_{-}^{B}(\omega)\right]+1-\eta,
\end{split}\\
\begin{split}
\mean{\hat{P}_{C}^{2}(\omega)}&=\left[\frac{3}{8}-\frac{1}{2\sqrt{2}}\cos(\omega\tau_{2})\right]\eta S_{-}^{C}(\omega)+\left[\frac{3}{8}+\frac{1}{2\sqrt{2}}\cos(\omega\tau_{2})\right]\eta S_{+}^{D}(\omega)\\
&+\frac{\eta}{8}\left[S_{-}^{A}(\omega)+S_{+}^{B}(\omega)\right]+1-\eta,
\end{split}\\
\begin{split}
\mean{\hat{X}_{D}^{2}(\omega)}&=\left[\frac{3}{8}+\frac{1}{2\sqrt{2}}\cos(\omega\tau_{2})\right]\eta S_{+}^{C}(\omega)+\left[\frac{3}{8}-\frac{1}{2\sqrt{2}}\cos(\omega\tau_{2})\right]\eta S_{-}^{D}(\omega)\\
&+\frac{\eta}{8}\left[S_{+}^{A}(\omega)+S_{-}^{B}(\omega)\right]+1-\eta,
\end{split}\\
\begin{split}
\mean{\hat{P}_{D}^{2}(\omega)}&=\left[\frac{3}{8}+\frac{1}{2\sqrt{2}}\cos(\omega\tau_{2})\right]\eta S_{-}^{C}(\omega)+\left[\frac{3}{8}-\frac{1}{2\sqrt{2}}\cos(\omega\tau_{2})\right]\eta S_{+}^{D}(\omega)\\
&+\frac{\eta}{8}\left[S_{-}^{A}(\omega)+S_{+}^{B}(\omega)\right]+1-\eta,
\end{split}
\end{align}
\end{subequations}
where $\tau_{1}$ and $\tau_{2}$ are corresponds to the time delay of the delay line $\Delta t$ and $N\Delta t$, respectively.

The oscillatory structure of the power spectrum is evidence that the phase locking of the system is correct and the period of the oscillation is determined by the length of each delay line. Moreover, since the spectra swap between $\hat{x}$ and $\hat{p}$ (eg. $\mean{\hat{X}_{A}^{2}(\omega)}$ is the same as $\mean{\hat{P}_{B}^{2}(\omega)}$), these observations ensure that we are indeed observing $\hat{x}$ and $\hat{p}$, and not the same quadrature. This cannot be distinguished from only the nullifier data.

The physical interpretation of frequency spectra can be explain as follows. In the case of an asymmetric Mach-Zehnder interferometer such as the one in one-dimensional cluster state, the optical delay is equivalent to a phase shifter whose phase is linearly dependent on frequency. Thus, the asymmetric Mach-Zehnder interferometer is equivalent to beamsplitters whose reflectivity depends on the frequency, which explains oscillatory behavior of the spectra. This behavior was observed in Ref. \cite{Yoshikawa2016}. In our two-dimensional case, the situation is also very similar; if we ignore BS-3, the experimental system is just two asymmetric Mach-Zehnder interferometer whose delay lines have different length. Therefore, when BS-3 is taken into an account, the spectra we observe will be equivalent to that of the one-dimensional case with 50\% loss on one arm. However, instead of contamination by vacuum in the case of the usual loss, one party of EPR pairs, which is equivalent to thermal noise, enter the state. The EPR noise is the term before the last term in Eq. (28). The first two terms are oscillations due to the Mach-Zehnder and the last term is the term due to optical losses.

\begin{figure}[ht]
\centering
\includegraphics[width=0.9\textwidth]{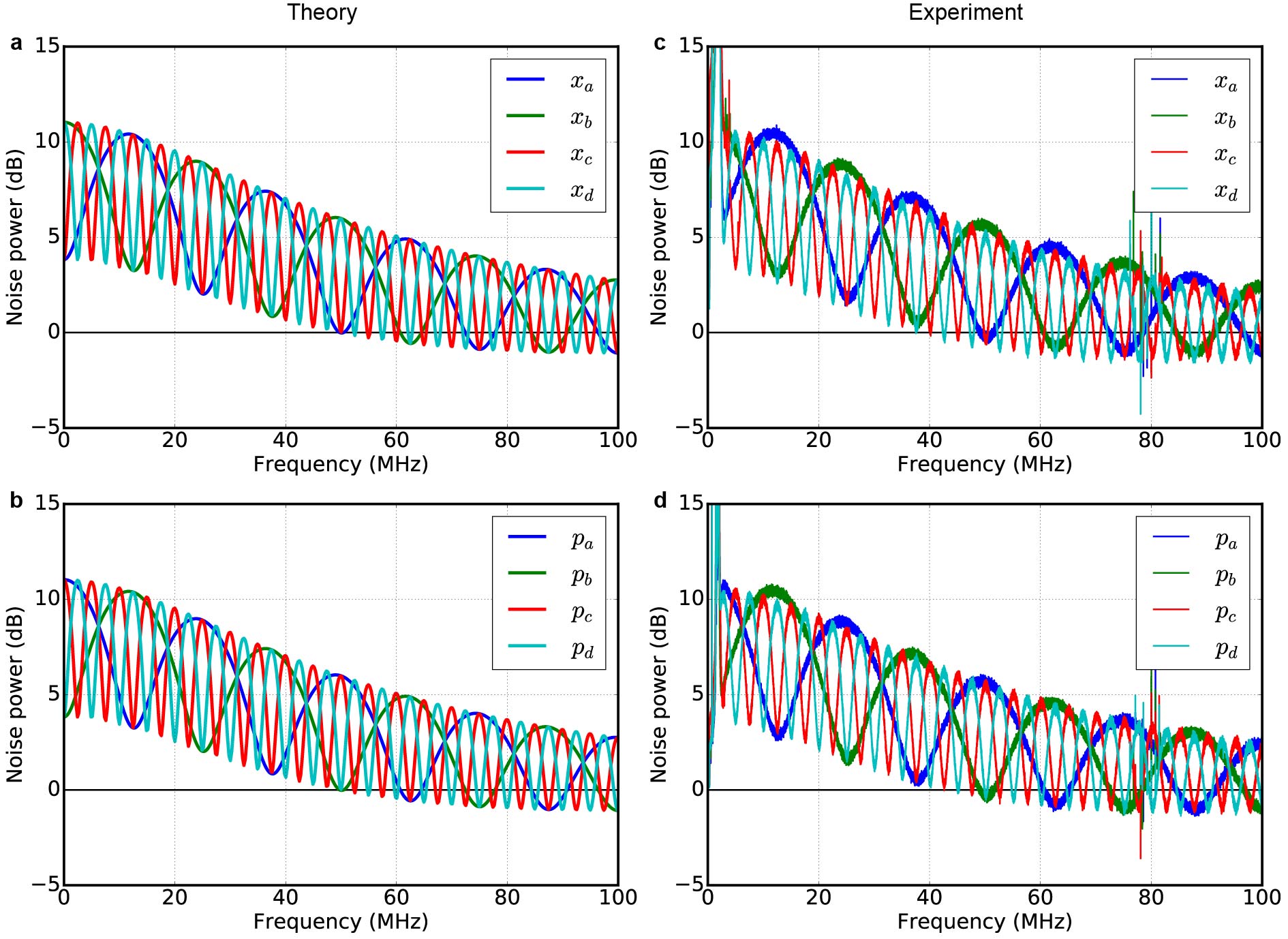}
\caption{Experimental measurement and theoretical prediction of power spectra of two-dimensional cluster state at each homodyne detector. (a) Theoretical prediction for $\hat{x}$. (b) Theoretical prediction for $\hat{p}$. (c) Experimental measurement of $\hat{x}$. (d) Experimental measurement of $\hat{p}$. For the theoretical plot, we assume constant $\eta=0.75$ and $\xi=0.65$.}
\label{fig:spectra}
\end{figure}

Figure \ref{fig:spectra} a,b show the theoretical prediction, while Fig.~\ref{fig:spectra} c,d show the Fourier transform of the electric signal obtained from each homodyne detector. The Fourier transform is performed using 524,288 data points obtained by recording electrical signal from homodyne detectors using oscilloscope with sampling rate of 1GS/s and an average of 3000 times was taken. We can see that the experimental results are in good agreement with the theoretical prediction and the shapes of the spectras are evidence that the phase locking of this system is correct. 

\section{Proof of Universality of Two-Dimensional Cluster State}
Finally, we will prove the universality of the generated state and discuss the actual usage in MBQC. In order to prove the universality, we have to show that our state can be used to implement arbitrary one-mode Gaussian operations, a two-mode Gaussian operation, and at least a single one-mode non-Gaussian operation. Also, when considering actual application in MBQC, it is desirable to program the operation easily and have every operation controlled independently.
\subsection{Measurement}
\begin{figure}[htb]
\centering
\includegraphics[width=\textwidth]{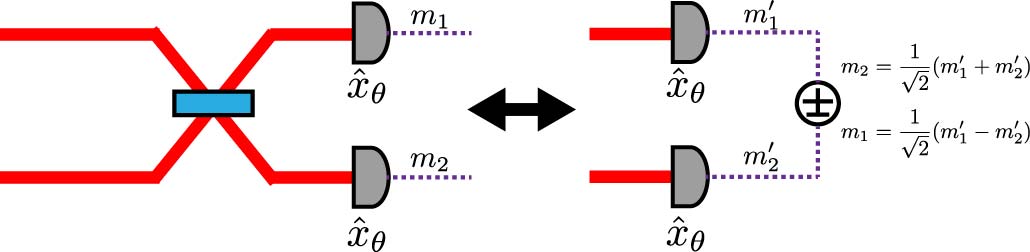}
\caption{Equivalence of measurement after beam splitter and measurement with post-processing}
\label{fig:clustermeasurement}
\end{figure}
\setcitestyle{super,open={},close={},citesep={,}}Let us first talk about homodyne measurements. As shown in Fig.~\ref{fig:clustermeasurement}, two homodyne measurements of same basis after beam splitter is the same as measuring the state without the beamsplitter and perform post-processing\cite{Alexander2018}. In the MBQC using our two-dimensional cluster state, we can perform such a measurement by using same measurement basis for homodyne-C,D and then performing post-processing of the measurement results. This property is also used to disentangle adjacent modes into separate distinct quantum wires.

\subsection{One-mode Gaussian Operations}
\begin{figure}[htbp]
\centering
\includegraphics[width=\textwidth]{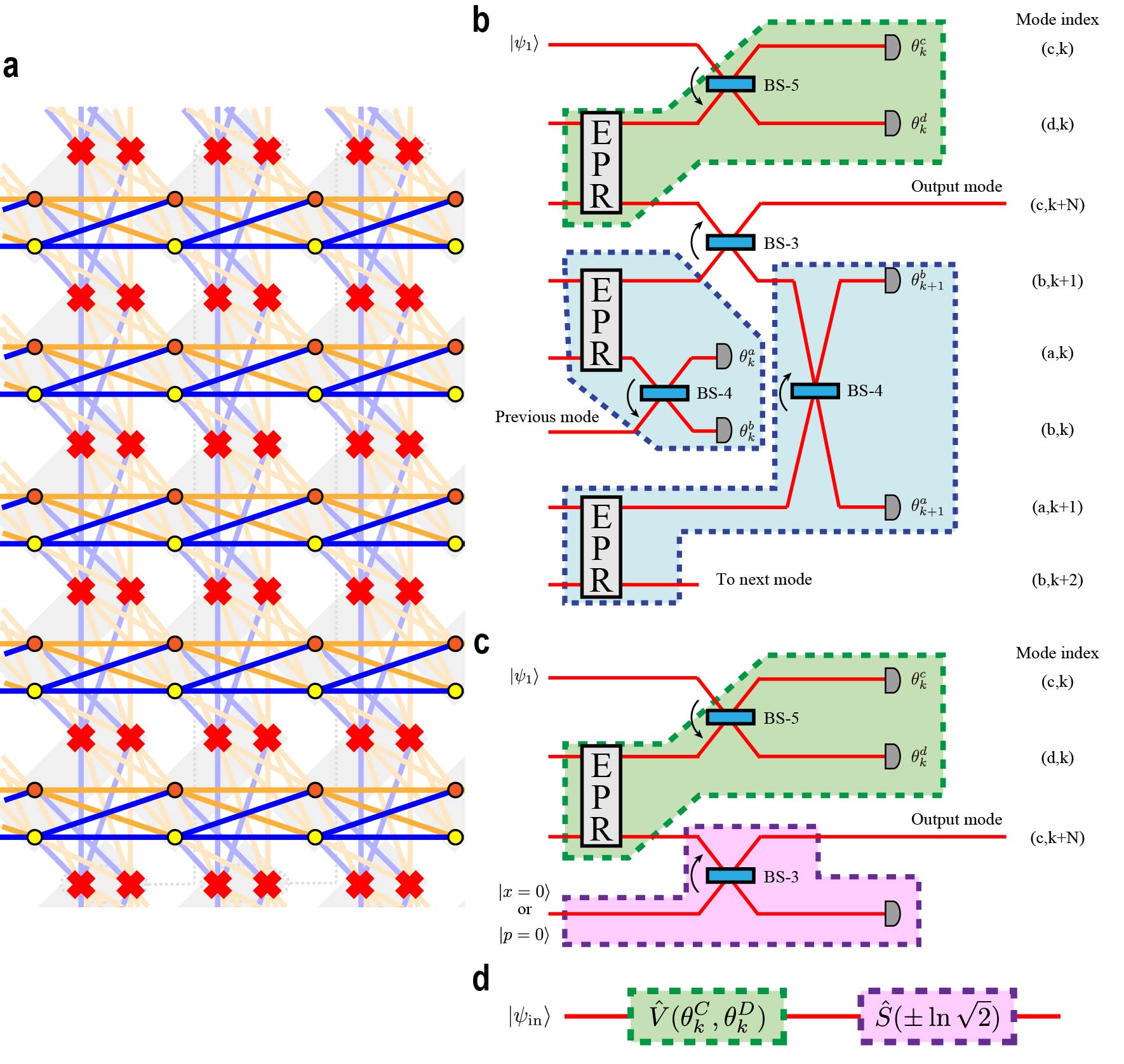}
\caption{Implementation of one-mode operation using this cluster state. (a) Choosing measurement basis so that adjacent modes are disentangled from each other. (b) Equivalent circuit for one step of one-mode operation. (c) Simplified circuit. (d) Total equivalent circuit. Green boxes and Blue boxes are quantum teleportation circuit with same correspondence as the box in the main text. Purple box is a universal squeezer.}
\label{fig:onemodeoperation}
\end{figure}
Recall from the main text that the computation using this two-dimensional cluster state can be represented by two types of circuits: one implements one-mode operations, while the other implements two-mode operations. Figure \ref{fig:onemodeoperation} shows how to implement one-mode Gaussian operations. To perform a one-mode operation, we have to disentangle the necessary part of the quantum wire from the adjacent quantum wires to prevent interactions. This can be done by selecting same measurement basis for quantum teleportation circuit in the blue box in Fig. \ref{fig:onemodeoperation}b. It was shown in Ref. \cite{Alexander2018} that by changing the measurement basis in quantum teleporatation circuit in the green box, we can implement the following one-mode operation,
\begin{align}
\hat{V}(\theta_{1},\theta_{2})=\hat{R}(\theta_{+})\hat{S}(\ln\tan\theta_{-})\hat{R}(\theta_{+}).
\end{align}
Where $\hat{R}(\theta)=\exp(-i\theta\hat{a}^{\dagger}\hat{a})$ is a rotation operation, $\hat{S}(r)=\exp\left[-i\frac{r}{2}(\hat{a}^{2}-\hat{a}^{\dagger 2})\right]$ is a squeezing operation, and $\theta_{\pm}=\frac{\theta_{1}\pm\theta_{2}}{2}$. After the green box, there are two blue boxes and a 50:50 beam splitter. Let us suppose we select $\hat{x}$ measurement basis for both homodyne detectors in the first blue box. This results in projection of the EPR state into $\ket{x=0}$ state when finite squeezing and displacements are ignored. Thus, if we measure $\hat{p}$ in the second blue box, we can consider the whole circuit after the first green box as a universal squeezer\cite{Filip2005} with $r=\ln\sqrt{2}$. Therefore, the whole operation can be written as
\begin{align}
\hat{U}(\theta_{k}^{C},\theta_{k}^{D})=\hat{S}(\pm\ln\sqrt{2})\hat{V}(\theta_{k}^{C},\theta_{k}^{D}),
\end{align}
where $\theta_{k}^{j}$ is the measurement basis of temporal mode with temporal index $k$ at spatial mode $j$. The resulting operation is equivalent to quantum teleporation plus squeezing. Note that because we have to restrict the measurement so that the adjacent modes do not interact, this implies that the measurement basis must satisfy $\theta_{k}^{A}=\theta_{k}^{B}$. When considering the universal squeezer, this implies further that the measurement basis for homodyne-A, B at the adjacent temporal index have to be orthogonal to each other and this results in the $\pm$ in the additional squeezing gate. Since there are two degree of freedom at homodyne-C, D, we can implement arbitrary one-mode Gaussian operations with two operational steps. Another way of looking at this additional squeezing gate is that when we perform one-mode operation, we are shaping the cluster state by erasing the part of the state which gives rise to entanglement between input modes. The effect of finite squeezing noise can be considered in the exact same manner as Ref. \cite{Alexander2014} which we will omit here.

\subsection{Two-mode Gaussian Operations}
\begin{figure}[htbp]
\centering
\includegraphics[width=\textwidth]{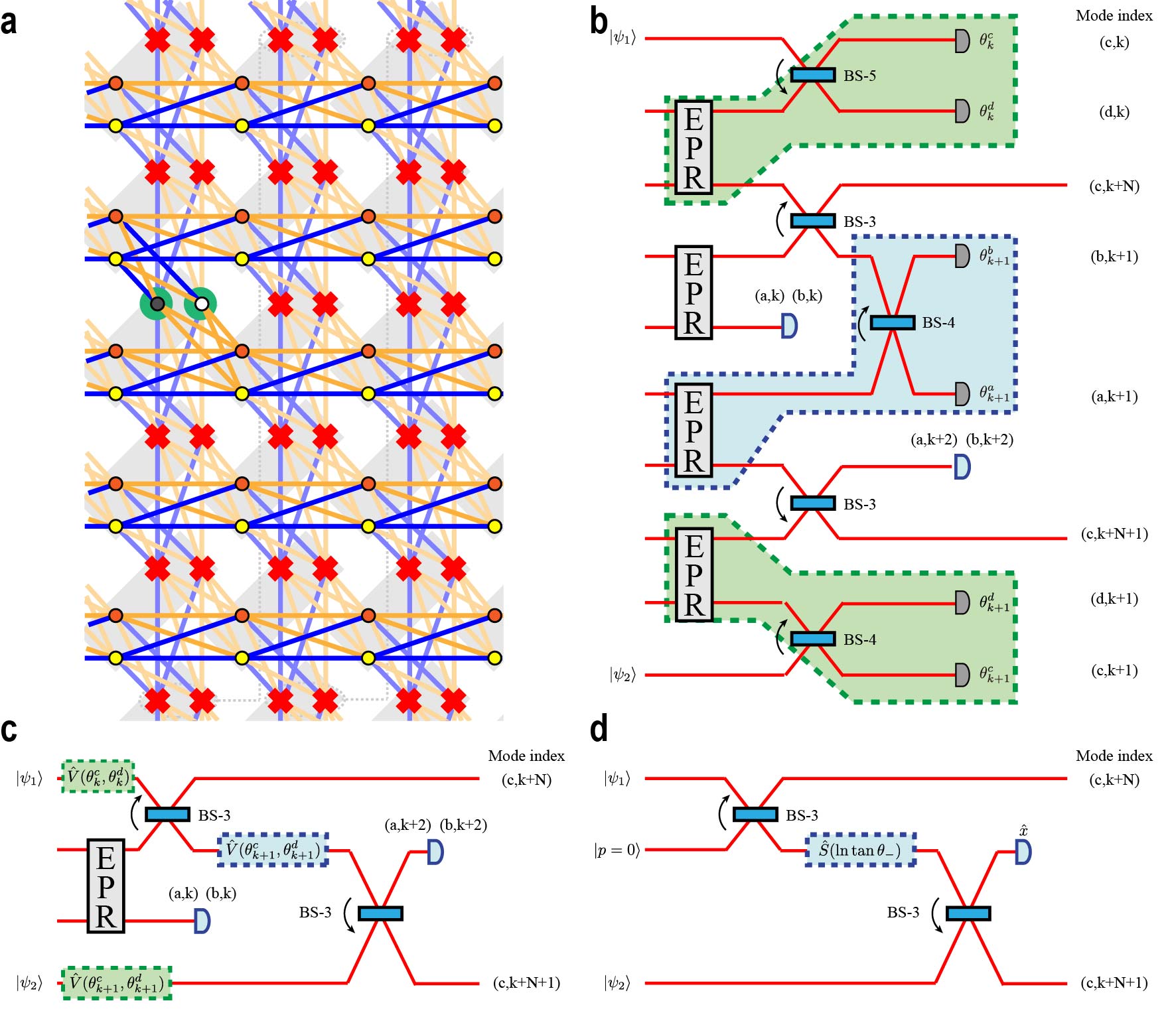}
\caption{Implementation of two-mode operation using this cluster state. (a) Choosing measurement basis to entangle adjacent input modes. (b) Equivalent circuit for one step of two-mode Gaussian operation (one-mode operation part is also included.) (c) Simplified circuit. (d) Example of implementation of quantum non-demolition interaction.}
\label{fig:twomodeoperation}
\end{figure}
To perform a two-mode operation, instead of using the same measurement basis for two homodyne measurements inside the blue box, we select a different measurement basis. Figure \ref{fig:twomodeoperation} (a) shows that by selecting the measurement basis properly, we can choose to do the two-mode operation between certain adjacent input modes. Fig.~\ref{fig:twomodeoperation}b show an equivalent circuit description. We isolate the subregion in the cluster state containing the two adjacent input modes on which we wish to perform the two-mode Gaussian operation from the rest of the cluster state by selecting measurements at $k$ and $k+2$, which are equivalent to the homodyne measurement denoted by blue detectors. Note that the actual measurement is done by two homodyne measurement and post-processing. By simplifying the quantum circuit, we arrive at equivalent circuit shown in Fig.~\ref{fig:twomodeoperation} (c). For example, if we project the EPR pair to $\ket{p=0}$, set the operation in blue box to be a squeezing operation, and measure $\hat{x}$ at the $k+2$ blue detector (Fig.~\ref{fig:twomodeoperation} (d)), we arrive at following input-output relation
\begin{align}
\hat{x}_{1,\textrm{out}}&=\sqrt{2}\left(\hat{x}_{\textrm{in},1}+\frac{1}{\sqrt{2}\tan\theta_{-}}\hat{x}_{\textrm{in},2}\right)\\
\hat{p}_{1,\textrm{out}}&=\frac{1}{\sqrt{2}}\hat{p}_{\textrm{in},1}\\
\hat{x}_{2,\textrm{out}}&=\sqrt{2}\hat{x}_{\textrm{in},2}\\
\hat{p}_{2,\textrm{out}}&=\frac{1}{\sqrt{2}}\left(\hat{p}_{\textrm{in},2}-\frac{1}{\sqrt{2}\tan\theta_{-}}\hat{p}_{\textrm{in},2}\right)
\end{align}
This corresponds to QND interaction $\hat{C}_{x}(g)=\exp(-ig\hat{p}_{1}\otimes\hat{x}_{2})$ with an additional squeezing gate on each mode. Thus, the operation in this case is
\begin{align}
\hat{U}_{\textrm{int}}(\theta_{k+1}^{A},\theta_{k+1}^{B})=\hat{S}_{1}(\ln\sqrt{2})\hat{S}_{2}(\ln\sqrt{2})\hat{C}_{x}\left(\frac{1}{\sqrt{2}\tan\theta_{-}}\right).
\end{align}

\subsection{One-mode Non-Gaussian Operations}
Now let us consider the implementation of non-Gaussian operation. As was suggested in Ref. \cite{Alexander2018} for the bilayer square lattice cluster state, non-Gaussian operations can be implemented with additional circuitry for implementing cubic phase gates located outside the cluster state. The cluster state generated in this work can also use the same method for implementing non-Gaussian operation, which completes the proof for universality. However, when we consider actual implementation, this method might not be the best method as it requires additional optical switch and homodyne detector. Also, while we can implement higher-order non-linear gate by combining multiple cubic phase gates, it is more advantageous to be able to perform higher-order non-linear gate directly. Here, we will describe two ways of implementing non-Gaussian operations. First, we show a method where we directly inject the ancillary state for cubic phase gate into cluster state. This method is applicable to arbitrary encoding for error correction. Second, if we use GKP encoding for error correction, then it was shown recently that distillable magic state can be generated by using only GKP qubits and Gaussian elements \setcitestyle{super,open={},close={},citesep={,}}\cite{Baragiola2019}. This method is very powerful in a sense that if we use GKP encoding, then the only necessary non-Gaussian component is GKP qubits.

\subsubsection{Cubic phase gate with ancillary state injection}
\begin{figure}[htb]
\centering
\includegraphics[width=\textwidth]{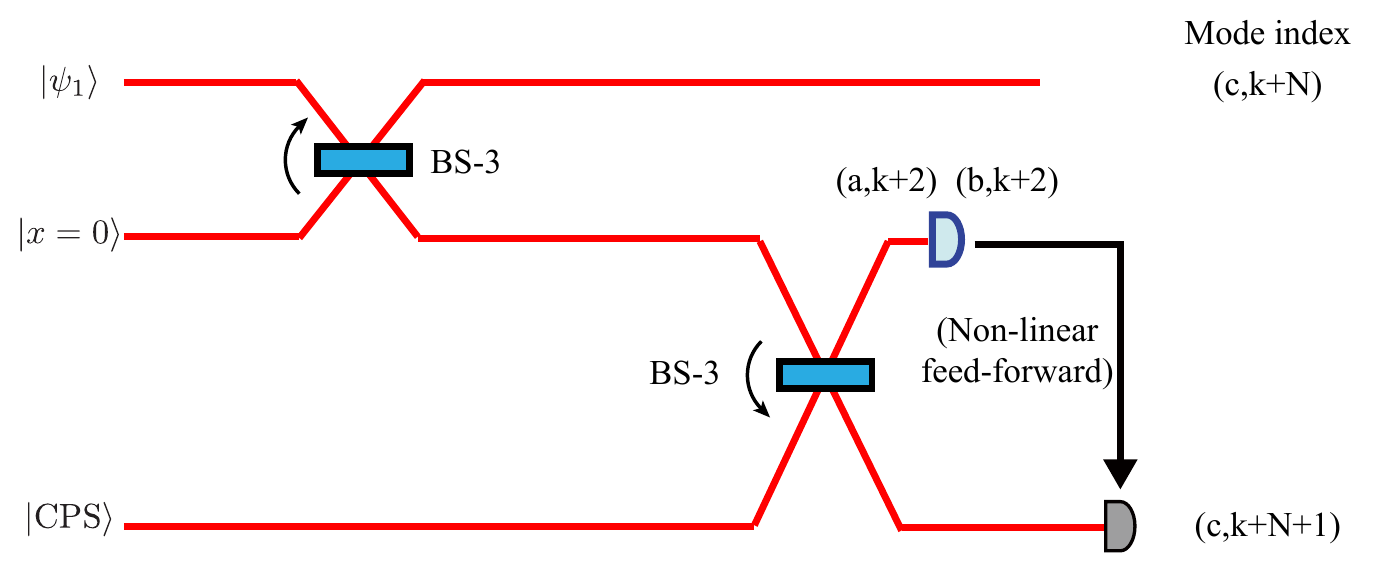}
\caption{Implementation of cubic phase gate by injection of cubic phase state into cluster state. This is the same circuit with two-mode operation where instead of two input modes, one of the input mode is cubic phase state (CPS) which is an ancillary state for cubic phase gate, and the homodyne measurements become adaptive where the measurement of mode $(c,k+N+1)$ depend on the results of the measurement at $(a,k+2)$ and $(b,k+2)$.}
\label{fig:nongaussoperation}
\end{figure}
By injecting the ancillary state for cubic phase gate into cluster state, it is possible to implement non-Gaussian operations for arbitrary encoding scheme without using any additional components. Also, in the light of recent discovery of method for non-linear measurements for higher order $\hat{x}^{n}$ gate\setcitestyle{super,open={},close={},citesep={,}}\cite{Marek2018}, injection of ancillary state into cluster state provides more flexibility and practicality compared to having to prepare additional circuitry for each non-Gaussian operation. One downside of this method is that part of the cluster state needs to be used for ancillary state injection, which reduces the number of input modes that the state can support. However, when we consider the scalability of this generation method, using some part of cluster state for ancillary state injection does not pose any problem to the computational capability.

To demonstrate this method, let us return to the circuit for implementing the two-mode operation. Instead of two input modes, let us make one mode cubic phase state\cite{Miyata2016}. In Fig.~\ref{fig:nongaussoperation}, when the measurement basis is properly chosen, this circuit is actually equivalent to the circuit for realizing cubic phase gate proposed in \setcitestyle{numbers,open={},close={},citesep={,}}Ref. \cite{Miyata2016}. This method requires adaptive homodyne, where the measurement of the one of the mode relies on the prior measurement result of the other mode. As retrieving quadrature values and calculating measurement basis for the second homodyne measurement requires time, in the experiment for cubic phase gate, additional optical delay line is required. However, as time-domain multiplexing already utilizes optical delays and the states are encoded in time, no additional delay line is required and the time available for computing in adaptive homodyne is determined by the length of the long delay line. Thus, implementation of cubic phase gate using our two-dimensional cluster state requires no additional physical resource which simplifies the actual implementation compared to the proposal in Ref. \cite{Alexander2018}. For higher-order non-linear phase gate, it has been shown that this can be implemented by tunable two-mode operation (either QND or beam splitter) together with injection of proper ancillary state and homodyne measurements. Since all of the aforementioned components can be realized with a cluster state, if sufficiently good ancillary states can be prepared, non-linear phase gate can be implemented using cluster states without additional quantum circuitry. 

\subsubsection{Generation of distillable magic state with GKP qubits}
Recently, it was also shown that if we use the GKP encoding to realize fault-tolerance, GKP qubits can also be utilized as distillable magic state for non-Gaussian operations. To do so, we only require GKP qubits, Gaussian operations and homodyne or heterodyne measurements. No photon counting or any other type of non-Gaussianity is required. This approach greatly reduces the necessary resource for both universality and fault-tolerance; namely, only GKP qubits are required for realizing both universality and fault-tolerance, in addition to two-dimensional cluster states. See Ref.~\setcitestyle{numbers,open={},close={},citesep={,}}\cite{Baragiola2019} for more information on this scheme.

\subsection{Feed-forward}
\begin{figure}[htbp]
\centering
\includegraphics[width=\textwidth]{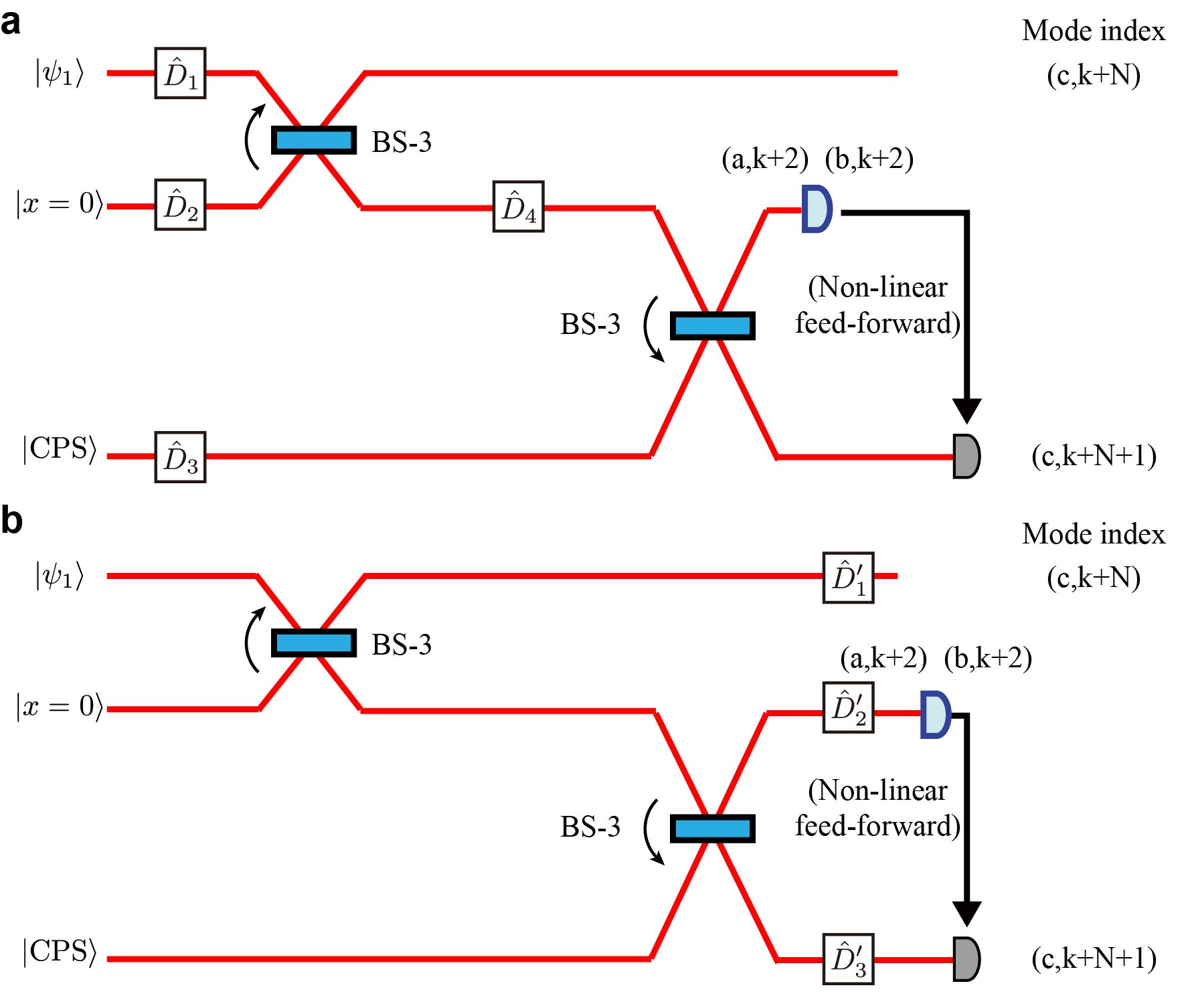}
\caption{Feed-forward operation for non-Gaussian operation. (a) Circuit showing all displacements from past operations. (b) Delaying of displacement through the network of beam splitters. The displacement before homodyne measurements can be absorbed into homodyne measurements. This method allows one to implement non-Gaussian operation without having to perform actual feed-forward operation in advance. Note that the resulting displacement ($\hat{D}^{\prime}$) depends on past measurement results and determines the quadrature values that needed to be subtracted when performing adaptive homodyne.}
\label{fig:feedforward}
\end{figure}
In the above discussion about implementing universal gates, we have omitted the discussion about feed-forward operations. In the usual implementation of MBQC, after the measurement, a corrective displacement operation must be implemented. If we restrict ourselves to multi-mode Gaussian operations, since displacement and Gaussian operations can be swapped at a cost of changing the amplitudes of the displacement operations, if we keep track of all measurement bases and results, the displacement operations can be delayed and implemented after all Gaussian operations are over. On the other hand, this is not possible with non-Gaussian operations. However, because the method for implementing non-linear quadrature phase gate proposed by Ref.~\setcitestyle{numbers,open={},close={},citesep={,}}\cite{Marek2018} consists of only Gaussian operation and adaptive homodyne measurements where the non-Gaussianity stems from the ancillary states, we do not have to perform the displacement operations before performing the non-Gaussian operation using this method; Displacements can be swapped with all the Gaussian operations and if we know the past measurement results we can take them into account when performing adaptive homodyne by just subtracting the value of the displacement from the homodyne measurement results (Fig.~\ref{fig:feedforward}). Furthermore, since we usually perform homodyne measurement on the output quantum state after implementation of quantum gates, we do not have to implement corrective displacement operations at all; the displacement can be taken into account by subtracting it from the measurement results of the final homodyne measurements. This means that the ability to adaptively change the measurement basis is the only feed-forward operation necessary. While these features are already present in Ref. \cite{Alexander2018}, we would like to emphasize that they also hold for the cluster state generated with the current setup.

\end{document}